\documentclass[11pt]{article}
\usepackage{amsfonts}
\usepackage{amsmath}
\usepackage{natbib}
\usepackage{hyperref}
\usepackage{graphicx}
\usepackage[margin=4.0cm]{geometry}
\usepackage{sidecap}
\usepackage{multirow}

\title{Populations in statistical genetic modelling and inference} 

\author{Daniel John Lawson\thanks{Heilbronn Institute for Mathematical Research, University of Bristol, Bristol, BS8 1TW, UK}}

\begin{document}
\maketitle

\section{Introduction}

It is human nature to look for patterns in the natural world.  One pattern is the grouping of individuals into populations based on observable features, be these anatomical, physiological, cultural, socio-economic, political or linguistic.  Many of these groupings are for convenience -- for example, we verbally group light frequencies into colours, and in hypothesis testing we define `case' vs `control' populations -- but some groupings are inherently meaningful.   
Variation within biological species has long been described by grouping individuals into `populations', though it is not clear \emph{a-priori} whether this is a correct description of genetic variation.
Focusing on sexually reproducing organisms, most individuals are genetically unique in some way, as typical organisms undergo numerous mutations per generation \cite{Lynch2010}, and sex acts to re-assort the genome of both parents \cite{CreightonMcClintock1931} leading to these mutations being found in complex patterns in later generations \cite{Hudson1983}.  Given that full pedigree\footnote{A pedigree is the full description of who was the parent of whom, until all individuals are traced back to the the most-recent common ancestor of the whole sample.} reconstruction is not possible in general, can we rigorously use the population concept to understand genetic variation? We argue that the answer is yes -- a very clear and unambiguous definition of a population is available using the well-understood statistical properties of inheritance. But also no -- applying this population concept to real organisms cannot be easily done in all cases.  Despite these limitations, the statistical concept of a population is extremely useful, both for conceptual understanding and detailed modelling of genetic variation.

There are two main approaches we can take to defining a population.  The first of these is a \emph{generative} modelling approach, in which populations are a component in an explicit model for how genetic variation arises. The second approach uses \emph{statistical} modelling, in which populations are inferred from genetic data. In both approaches, there is in fact a generative model for the genetic structure within a population -- they differ on how differences between populations are treated.  A population can be defined statistically as a group of individuals that cannot be distinguished on the available data, a less precise definition than is needed when using generative models.  Under certain ideal -- but reasonable -- conditions that we discuss below, these two definitions overlap, allowing a very clear definition of what we mean by `population'.  Under this framework, current tools allow for significant progress in understanding many features of genetic structure.  Of course, reality can be more complicated and there are at least some cases for which the specifics of genetic inheritance and demography can make even sophisticated attempts at classification into populations seem clumsy and arbitrary. 

Central to the majority of theoretical approaches is the idea that, over relevant timescales, all individuals within a single population can be viewed as mating at random within that population (see e.g. \cite{Wright1984}).  This timescale caveat is extremely important, as we know that real individuals of any species do not randomly mate. Yet mating decisions -- based on chance factors such as geographical location and certain individual preferences -- are, in most circumstances, `self-averaging'\footnote{A self averaging process \cite{Lifshitz1952,KadanoffStatPhys2000} is one for which averaging over a single realisation results in the same answer as averaging over many realisations, and occurs most simply when random fluctuations are \emph{additive}, i.e. not state-dependent.} such that over many generations, random mating is an appropriate approximation.  Random mating is a useful assumption as it describes the simplest unit of population structure, on which complexity can be built.  But -- particularly with modern advances in data acquisition \cite{Schuster2008,Mardis2011} -- detection of structure within populations is possible \cite{Lawson2012}.  Were sufficiently detailed data available, all individuals would form their own `population'.  This occurs because even though we cannot infer the full pedigree, we can still detect the presence of familial relationships \cite{Pemberton2010,Kyria2011,Stevens2011} such as distant cousins, which violates the random mating assumption. We will discuss relaxations to allow for this type of relatedness, which requires an appropriate description of the sort of deviation we expect within a population.

Population concepts transcend species boundaries and we aim to be as general as possible.  The random mating assumption is the starting point for the theoretical justification of the approach, and applies with varying degrees of accuracy depending on the genetic and life cycle details of organisms.  Generally speaking, sexual species are well modelled by this approach, varying from a near-perfect description of Drosophila \cite{HendersonLambert1982}, an adequate description for humans (e.g. \cite{Rosenberg2002}, see below), to a difficult fit for fungi \cite{Milgroom1996} and plants \cite{Platt2010}.

Not all species reproduce sexually.  Sexual species (including most animals) have two (or possibly more in the case of plants) copies of each chromosome, one of which is inherited from each parent.  In contrast, single-celled organisms (prokaryotes such as Bacteria and Archaea) have a very different genetic structure with only a single copy of each chromosome.  Standard sexual reproduction is therefore not possible and it might be assumed that random mating is not an appropriate description at all.  However, many such species can be represented using this approach if their recombination rate is sufficiently high \cite{Falush2003} -- Bacteria can still exchange DNA via non-sexual reproduction, by taking a small fragment of DNA from another organism either accidentally from the environment or via various deliberate (evolved) mechanisms.  The amount of recombination bacteria engage in is highly species dependent.

When recombination is rare or absent, a different population concept is needed.  Individuals that occupy the same ecological niche compete neutrally, in which case individuals are sampled under an important genetic distribution called the `coalescent' \cite{Kingman1982}, discussed in detail below.  The genetic distribution within a non-recombining population is very different to that for a recombining population. Rather than variation around a single genetic type being associable with a population, very distinct genetic subtypes can arise due to random fluctuations (since individuals do not exchange genetic information). The coalescent fully describes how genetic variation is distributed, but this typically splits up into multiple separate sub-types when variation is viewed in few dimensions \cite{LawsonJensen2007}.  Therefore, for non-recombining organisms, a population concept exists as more of an ecological constraint on evolution.  Although this population concept is interesting, it is very different from that which is seen in our focus of recombining organisms.

Data collection is predominantly based around sequencing or genotyping many individuals from a small set of sample locations (e.g. \cite{Rosenberg2002,Hapmap2007}).  This naturally leads to finding population-like structures in the data because individuals found at the same location tend to have similar genetic history.  However, as sampling becomes more dense, and genomic techniques more precise, a more continuous picture arises.  Strong substructure can be found within single sampling locations (e.g. \cite{Lawson2012}) and yet no clear demarcation may exist between samples.  Whilst the concept of a randomly mating population is helpful for understanding the genetic structure, we will be increasingly required to work with more complex descriptions.  Although there is no clear solution to these issues, we will discuss some possible approaches and how they relate to standard methods.

\section{Generative models} \label{sec:gen}

Mathematical or computer models that can simulate artificial datasets are called generative models.  The concept of a population, and the generation of the features of the population, are explicitly defined in the model and individuals can be sampled from them. Generative models typically allow for very complex features to be included, but are hard to perform inference with.  The concept of a population used by such models can be very flexible and cannot be generally defined.  Typically however, populations appear as randomly mating groups of individuals, with additional and often complex modelling for how individuals in different populations can interact.

Many population modelling approaches involve generative models, and we focus on two strands here.  The first grew from `statistical phylogeography' which used `multiple locus coalescent models' \cite{KnowlesMaddison2002,Chikhi2002} that consider several `independent' genomic regions (loci) within which recombination is absent.  On a single locus, a `population' has the same meaning as for non-recombining species described above.  Loci are then combined assuming a randomly mating population. Modern incarnations (e.g. \cite{Thornton2006}) consider fully sexual populations in a historical scenario in which an ancestral population experiences separation and subsequent genetic drift (i.e. populations are related via a tree, possibly with migration).  Throughout this history, populations randomly mate.  Inference of the full demographic history requires complex inference tools such as Approximate Bayesian Computation \cite{Beaumont2010}.  However, some historical hypotheses, such as the presence of gene-flow between populations (i.e. admixture), can be tested for directly (e.g. \cite{Durand2011}) in certain restricted ways.  Model organisms such as the fruit fly drosophila \cite{Thornton2006} and humans (e.g. \cite{Cornuet2008} and references within) have been treated to the full modelling approach.  However, the multiple locus approaches are widely applicable to non-model organisms \cite{Knowles2009} from plants \cite{HerreraBazaga2008} to frogs \cite{Bonin2006} as costly data collection is not needed.

\begin{figure} [ht]
\begin{centering}
\renewcommand{\thempfootnote}{\fnsymbol{mpfootnote}}
\includegraphics[clip,width=0.9\textwidth,angle=0]{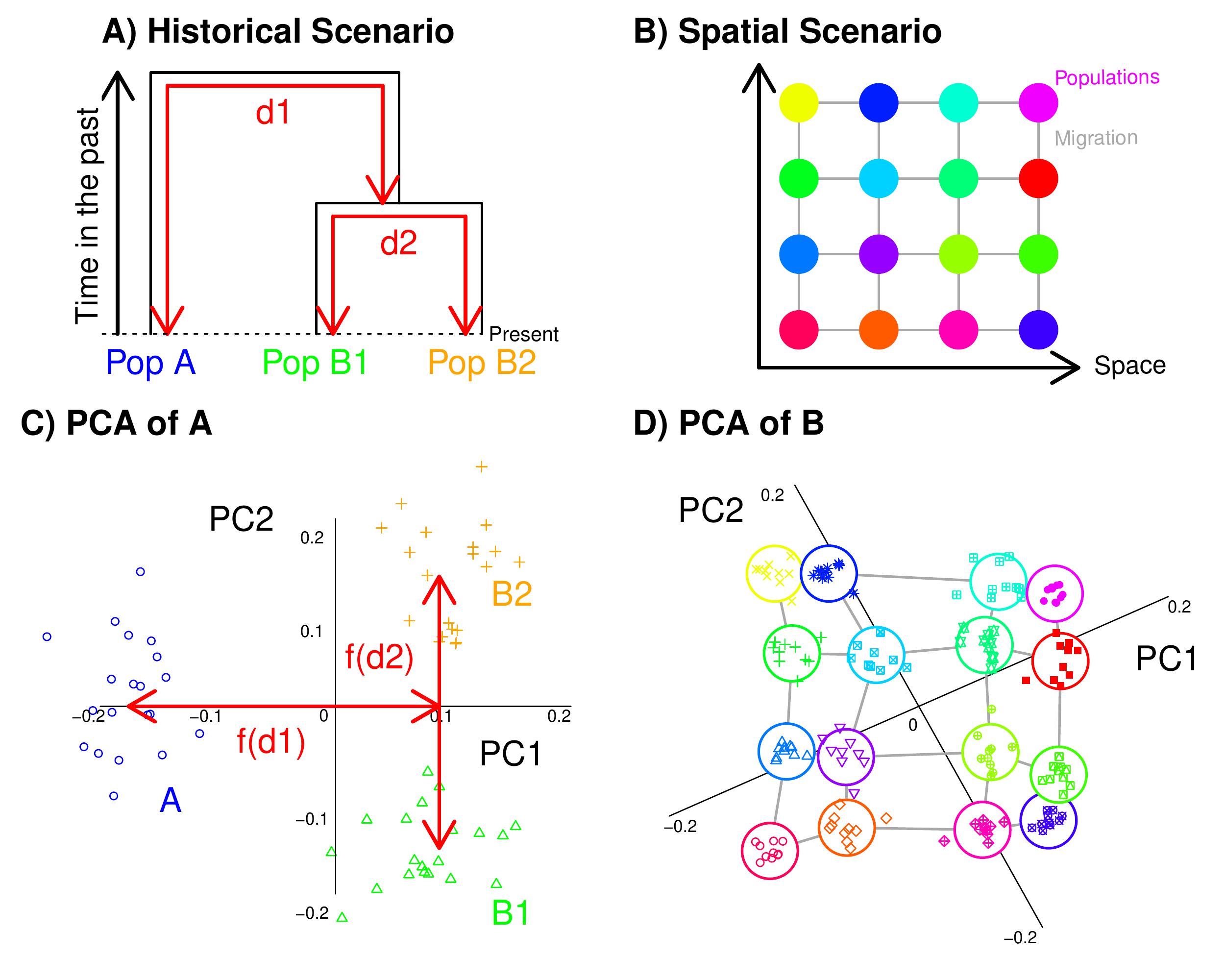}
\caption[Simple Population Models]{Simple population scenarios*
and the resulting PC decomposition.  A) A Simple hierarchical tree model for population splitting.  d1 and d2 represent genetic drift times.   B) A deme-based spatial migration scenario in which migration occurs at constant rate along edges and each node contains a population.  C) The PC decomposition of the scenario in A. Each PC is associated with an independent drift component, with genetic drift time affecting both the Eigenvalue, and the distance relative to the within-population variation (i.e. f(d1) and f(d2)). D) the PC decomposition of C recapturing the spatial structure in this case.  Were sample sizes not even, this may not have occurred, and higher PCs contain a complex pattern created by local drift and historical admixture.

*\small{\emph{For A, 1000 unlinked SNPs simulated as in \cite{Lawson2012}.  For B, Simulated using the software msms \cite{EwingHermisson2010} sampling 20 individuals per population with the parameters `-r 500 -t 50' resulting in 6754 linked SNPs, and recent migrants removed to reduce the dataset to 10 individuals per population.}
\label{fig:fig1}}}
\end{centering}
\end{figure}

A lot is understood about the genetic structure produced under a tree model for population history, i.e. if individuals mate randomly within populations that have been subjected to arbitrary genetic drift without admixture.  The observed genetic variation typically takes the form of Single Nucleotide Polymorphisms (SNPs, which are 0 or 1, either present or not).  While the raw distribution is difficult to visualise, differences between individuals (and their populations) can still be plotted.  \cite{McVean09} describes how simple generative models project into their Principle Components (PCs) based on decomposing the genetic structure into independent directions of variation\footnote{We use a Singular Value Decomposition (SVD), a generalisation of PC analysis.  The SNP data matrix $X$ containing $l$ rows (SNPs) and $n$ columns (individuals) is decomposed into $U D V^T$ where $U$ is an $l$ by $n$ matrix, $D$ is a rectangular diagonal matrix of the eigenvalues, and $V$ is the $n$ by $n$ eigenvector matrix relating individuals. Columns of $V$ are the PCs (which are usually what is plotted), and each row is an individual. \cite{LawsonFalush2012} has more details about the interpretation.}.  These directions are ordered so that the direction of largest variation appears first; see Figure \ref{fig:fig1} C-D.
Individuals that share many SNPs tend to be close, and SNPs that vary consistently between individuals tend to dominate the higher PCs (those that are typically plotted).  The lower PCs describe SNPs that show no particular pattern and are usually considered to be `noise', e.g. random sampling. Individuals within a single population are characterised by having the same SNP frequencies, and hence approximately share the same location in PC space. 

Differences between populations in a PC analysis can be interpreted.  Genetic drift affects the PCs by separating the mean position of populations, as shown in Figure \ref{fig:fig1} A,C with distances $d1$ and $d2$ becoming separations $f(d1)$ and $f(d2)$.  Additionally, an individual who is admixed between two populations is located between them in PC space (Section \ref{sec:admix}).  Unfortunately, sample size strongly affects both the location and order of drift effects in the PCs, and interacts with both drift and admixture in a non-trivial way to determine the location of populations in PC space. Even worse, although an admixture scenario has a predictable effect on the PCs, this is not always inferable (see Section \ref{sec:stat}).

Another generative model is demic diffusion \cite{AmmermanCavalli-Sforza1984,Sokal1991} 
which captures complex geographical structure.  \cite{Waples2006} perform a thorough review of this approach, with specific emphasis on what a population means in this context.  Again, populations are randomly mating groups of individuals living in demes at a specific geographic location, and migration occurs between nearby demes (see Figure \ref{fig:fig1} B,D). Like the explicit historical model above, complex histories can be specified.  However, the full history of a population is only implicitly described in this approach, as individuals migrate between demes.  Recent migrations (e.g. within the last few generations) do not fit neatly into a population description.

Demic diffusion models and PC analysis are usually used together to make inference about the nature of historical and geographical processes shaping evolution.  Unfortunately, there are many historical scenarios consistent with the major PCs \cite{Francois2010} which casts doubt on the inferences that can be directly made.  In principle, there is more information contained in the lower PCs, describing e.g. genetic drift unique to each population, but this is difficult to extract.  
This is in contrast to simple statistical populations (Section \ref{sec:stat}) which can be inferred using PC analysis \cite{LawsonFalush2012}.  The inference problems are related to the presence of strong admixture in these models, which is discussed in Section \ref{sec:admix}.  The utility of PC analysis for inference about population structure is therefore a contentious topic.

There is more that can be said about PC analysis; in particular, methods accounting for linkage (Section \ref{sec:stat},  details in \cite{LawsonFalush2012} and  \cite{Lawson2012}) can be used to improve PC decomposition.  However, complex  genomic scale datasets are not completely described by the top few PCs and so we view PC analysis as a useful visualisation tool rather than a primary analysis technique for understanding populations.

\section{Statistical definitions of a population} \label{sec:stat}

Statistical approaches such as STRUCTURE \cite{Pritchard2000} define a population as a cluster of statistically equivalent individuals.  For SNP data taking the value of either 0 or 1, this means that there is a single frequency for each SNP within each population, and a population is defined by the full set of SNP frequencies.  Although a single SNP is not very informative (unless it perfectly segregates populations), multiple SNPs can be averaged to identify fine population differences.

Each SNP in an Individual in the STRUCTURE model is sampled `independently and identically distributed' (IID) from the population SNP frequency.  This means that SNPs are both independent, and are all drawn with the same frequency, so there can be no correlation between the SNPs from e.g. related individuals.  More generally, we can define a population as a collection of individual who are genetically \emph{exchangeable} \cite{Bernardo1996}.  Exchangeability within a population is less restrictive than IID sampling -- the SNPs between individuals may be correlated, providing that the order that individuals are placed within the population does not matter\footnote{A classical example is the P\'{o}lya urn \cite{JohnsonKotz1977} in which a red and blue ball are placed in an urn, and then a ball is repeatedly removed at random and replaced by two balls of the same colour.  The probability of drawing a red ball depends on the history of the process but knowing only the number of each ball in the Urn is enough - the order does not matter.  The coalescent process in genetics \cite{Kingman1982} is but one very important example of exchangeability in action.}.

In the STRUCTURE model, related individuals are assigned their own population which has a different SNP frequency from the population the individuals were sampled from (and may be close to 0 or 1).  More general definitions of a population based around exchangeability
would allow individuals to be sampled as e.g. a mixture between a random individual (i.e. using the population frequency) and a previously sampled individual. The number of related individuals within a population and the degree of relatedness would be additional parameters to be inferred.  Such models are more complex to implement (and hence are not considered in any major inference algorithm), but should remain part of our conceptual understanding of a population.

\begin{figure}[!ht]
\begin{minipage}{\textwidth}
\begin{centering}
\includegraphics[clip,width=1.0\textwidth,angle=0]{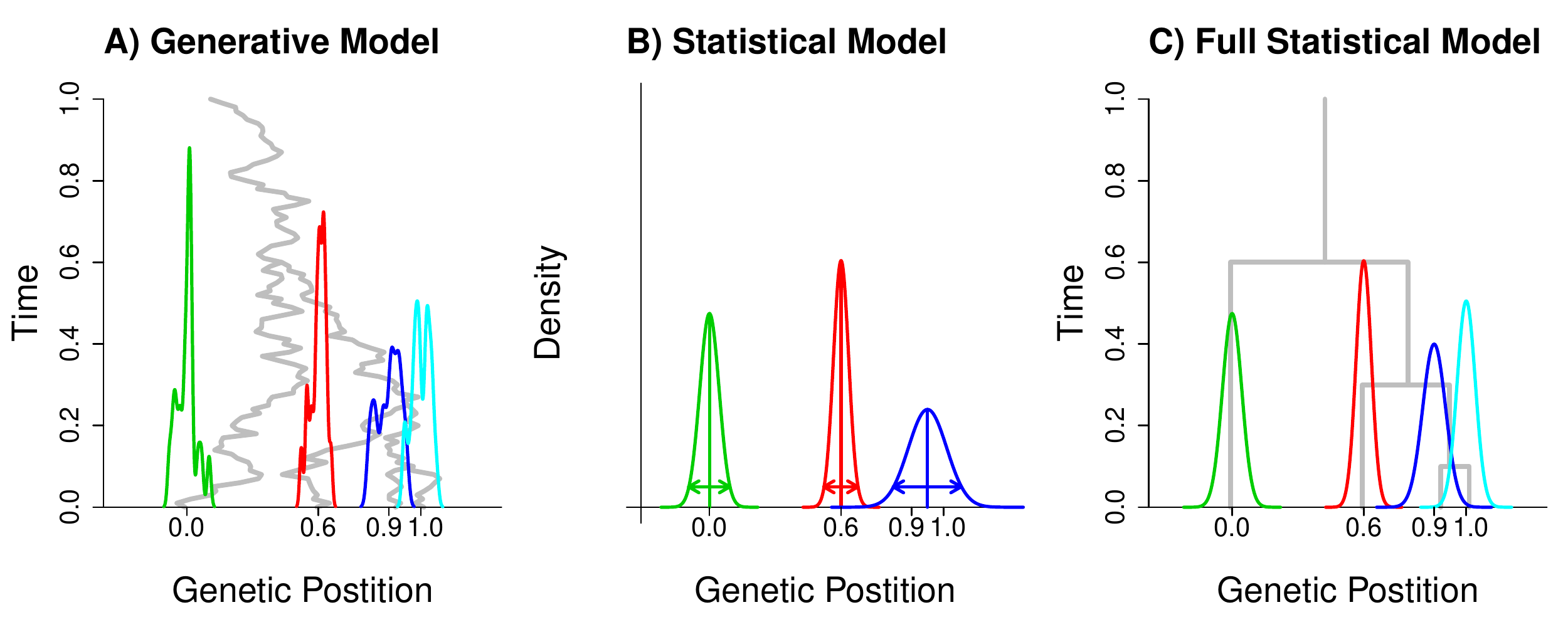}
\caption[Statistical Model Types]{Populations as they may hypothetically appear in different types of model. A) A fully generative historical model describing the genetic drift process (grey tree) differentiating populations (`peaked' curves centred on marked locations on the x-axis) in genetic space (i.e. a direction in PC space). The details of the drift process must be specified.  B) A simple statistical model, in which only the within-population model (here the means and variances) must be correctly described, and population-level parameters are inferred. Populations may be merged (right-most populations) if they are not sufficiently distinct. C) A sufficiently flexible statistical model for the population-level parameters can allow these to be inferred more accurately; here a hierarchical model matches the data better and therefore infers the subtle population differences correctly.
\label{fig:fig2}}
\end{centering}
\end{minipage}
\end{figure}

To say that individuals belong to different populations, we only need a model of how individuals within a single population vary, as STRUCTURE does via the SNP frequency.  In a simple statistical approach, a correct model for how differences between populations arise is not needed (Figure \ref{fig:fig2}B). This contrasts with generative models (Figure \ref{fig:fig2}A) which require an explicit description of how the current genetic distribution was reached.  Despite this, a model for population differences is very helpful as it increases statistical power to separate populations (Figure \ref{fig:fig2}C). In this example, the underlying generative model (Figure \ref{fig:fig2}A) is diffusion along a tree.  Because the underlying population separation method was hierarchical, a tree-based statistical model can capture the main differences between populations.  Models of population differences are sometimes omitted because a) sufficiently flexible models should all tend to the truth as the amount of data gets large, even if they ignore between-population differences, and b) if the model is incorrect, it can lead to problems.

\begin{figure}[!htbp]
\begin{centering}
\makebox[1.1\textwidth]{
\hspace{-1.5cm}
\includegraphics[clip,width=1.05\textwidth,angle=0]{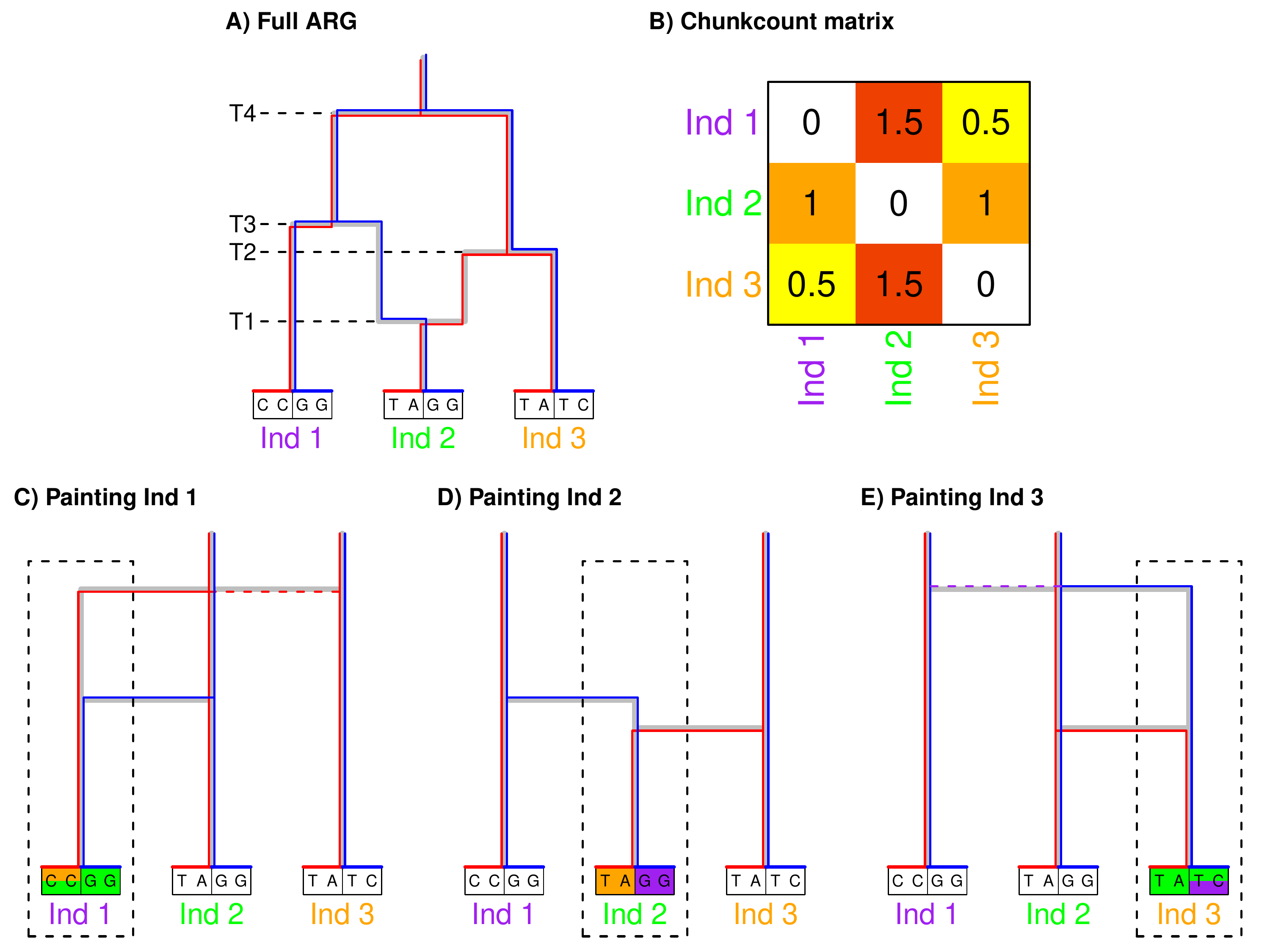}
}
\caption[ChromoPainter Illustration]{Cartoon illustrating the ChromoPainter model as an approximation to the ARG on a simple sequence.  A) The ARG describes the full history of recombination and coalescence between individuals, terminating with a Most Recent Common Ancestor.  The ARG consists of a series of coalescence events (times T2--4), and recombination events (time T1).  Recombinations split the genome into segments that have (possibly) distinct tree relationships between individuals (red and blue trees, in the genomic regions shown by the bars).  B) The ChromoPainter chunkcount matrix counting the number of haplotype chunks shared between each pair of individuals (which is asymmetric in general).  It counts the \emph{expected} number of haplotypes blocks used by the painting (C-E), rather than the number of SNPs.
C-E) The chunkcount matrix approximates the ARG by describing only the most recent relationship for each individual separately at each genomic location.  Coalescence and recombination events are forbidden between other individuals.  Where multiple individuals are equally close (dashed lines in C and E), these are averaged and fractions of the chunk are assigned from each. For example, In Painting Ind 1 (Figure C), SNPs 1-2 are on the same haplotype which is uncertain between Inds 2-3, using 0.5 chunks from each.  Due to requiring fewer mutations, SNPs 3-4 can be inferred as from Ind 2 (using a single chunk), leading to the top row of (B).
\label{fig:fig3}}
\end{centering}
\end{figure}

The genetic structure within a randomly mating population is well understood, and is characterised by the mathematical model known the Ancestral Recombination Graph (ARG) \cite{Griffiths1981,Hudson1983} (Figure \ref{fig:fig3}A).  This describes the genetic history of all individuals using three basic ingredients. Firstly, exactly when their ancestors were shared with ancestors of other individuals in the sample (coalescent events; times T2-4). Secondly, when ancestors mated with ancestors of other individuals (i.e. recombination events which split the genome between these two parents, time T1). Thirdly, when mutations arose (although for our purposes mutation times need not be inferred as these can be integrated out). As such, the ARG is a complete probability model of a closed population\footnote{The ARG specifies the probability of a particular set of recombination and coalescent events. We start with $k=N$ lineages, one for each genetic sample. Coalescence between lineages occurs at rate $1$ between all $k(k-1)/2$ pairs of lineages, and recombination occurs on each lineage at rate $\rho$.  Therefore the time between events is exponential with rate $k(k-1)/2+k\rho$. The coalescent model is the ARG with $\rho=0$.}.  Note that an ARG without recombination reduces to the coalescent model, which is a tree with a probability model for the coalescent times.  Although the ARG is a statistical model, it is equivalent to a generative model of a neutrally competing population with fixed recombination rate, mutation rate, and population size \cite{Hein2005}, and a wide variety of different modelling assumptions all lead to the ARG under reasonable limiting conditions, such as large population sizes so that mating can be considered to occur randomly.

Unfortunately, it is not practical to perform inference with the ARG, and so most approaches are based around approximations to it.  Almost all population inference methods start by assuming that genetic drift is weak, because this is relatively harmless as we can always detect the difference between populations when drift is strong.  Another common approximation is that the recombination rate is high enough to treat SNPs independently.  The most widely used tools, including STRUCTURE \cite{Pritchard2000}, STRUCTURAMA \cite{Huelsenbeck2007}, PARTITION \cite{Dawson2001} and ADMIXTURE \cite{Alexander09} start with this assumption, which leads to modelling SNPs only via their SNP frequency as described above.  

The SNP frequency approach works well when SNPs are approximately independent, but becomes increasingly inaccurate in the presence of linkage disequilibrium (LD).  LD is the correlation between SNPs along the genome due to sharing the same genetic history; for example, in Figure \ref{fig:fig3}A, SNPs 1-2 are in LD because they are each described by similar (and in this case, the same) trees.  As more recombination events split the genome between two SNPs they become less correlated.  Therefore, LD is stronger in genomic-scale sequencing datasets produced by modern sequencing technologies than in older techniques that sampled fewer loci.  Modelling LD is becoming more important, and although STRUCTURE has an option \cite{Falush2003} for this, the approach is too computationally costly for most applications.

A powerful approach for handling LD, which has an additional benefit of speeding the computation, is that of ChromoPainter \cite{Lawson2012}.  This uses the `painting' algorithm of \cite{LiStephens2003} to identify long tracts of genetic material that have been inherited as a single contiguous chunk, which can be seen as an approximation to the ARG (Figure \ref{fig:fig3}).  Each haplotype\footnote{A haplotype is a particular copy of a chromosome; recall that humans and other animals have two copies of each chromosome, which are called haplotypes. Current sequencing technology cannot tell which copy of DNA a particular mutation is on, and therefore we observed the `genotype', i.e. the mixture of the two.  Haplotypes are inferred using a technique called phasing to separate out which mutations appear on which copy of the chromosome.} is described at each genomic location by finding the haplotype from a pool of potential donors that most closely matches it.  The donor at a particular genomic location has to be inferred, and there is a constant probability per unit recombination distance\footnote{Recombination does not occur uniformly over the genome but most frequently occurs at `recombination hotspots' \cite{Myers2010}.  This can be accounted for in the model by measuring distance along the genome in recombination distance.} that the donor will change to any of the other possible donors.   The result is that `chunks' of DNA are found in which relatively few SNPs differ between the donor and the haplotype being painted.  The model takes a computationally convenient form called a Hidden Markov Model\footnote{In our context, the presence of a mutation at a genomic location $i$ depends only on who the donor haplotype $x_i$ is (which is `hidden').  In turn, the probability of the donor haplotype $x_{i}$ depends only on the donor haplotype at the previous location $x_{i-1}$ and not on any other genomic positions (this is what is meant by `Markov').}, which allows for a fast inference of a wide range of useful summaries with complete accounting for statistical uncertainty.  We focus on the `chunk count' matrix $X_{ij}$ -- the (expected) number of chunks donated to individual $i$ from individual $j$.

Under reasonable assumptions (details in \cite{Lawson2012}), a shared haplotype chunk can be viewed as a single observation of coinheritance between two populations in a similar way to the sharing of a SNP.  The FineSTRUCTURE model combines this information over all individuals by treating a population as having a characteristic rate of donating chunks to, and receiving from, each other population. The only information required is the number of chunks donated between each pair of individuals\footnote{This is a sufficient statistic for the model likelihood.}, which is a matrix of dimension $N \times N$ -- a significant reduction in the amount of data that must be processed as there is no dependence on the number of SNPs.  Additionally, there are no population-level parameters to be inferred, which makes inference both very fast and very reliable, as the parameter sampling does not tend to get stuck in local modes. SNP frequency based approaches (e.g. STRUCTURE and ADMIXTURE) must infer the underlying SNP frequencies which can be multimodal.  FineSTRUCTURE can therefore handle large sample sizes; thousands using MCMC\footnote{Markov-Chain Monte Carlo (e.g. \cite{Gamerman1997}) is a very general and powerful inference technique based around sampling parameters near to previously sampled parameters, as these are likely to be `good' at describing the data (i.e. have a high likelihood).  The MCMC samples converge to independent samples from the Posterior distribution if the sampler is run for `long enough', which may take a long time.}, and potentially many more using greedy optimisation approaches. However, whilst fast, greedy methods usually do not find the best solution and may get stuck at sub-optimal parameter values that cannot be improved by small changes.

\begin{figure}[!phtb]
\begin{centering}
\vspace{-2cm}
\includegraphics[clip,width=\textwidth,angle=0]{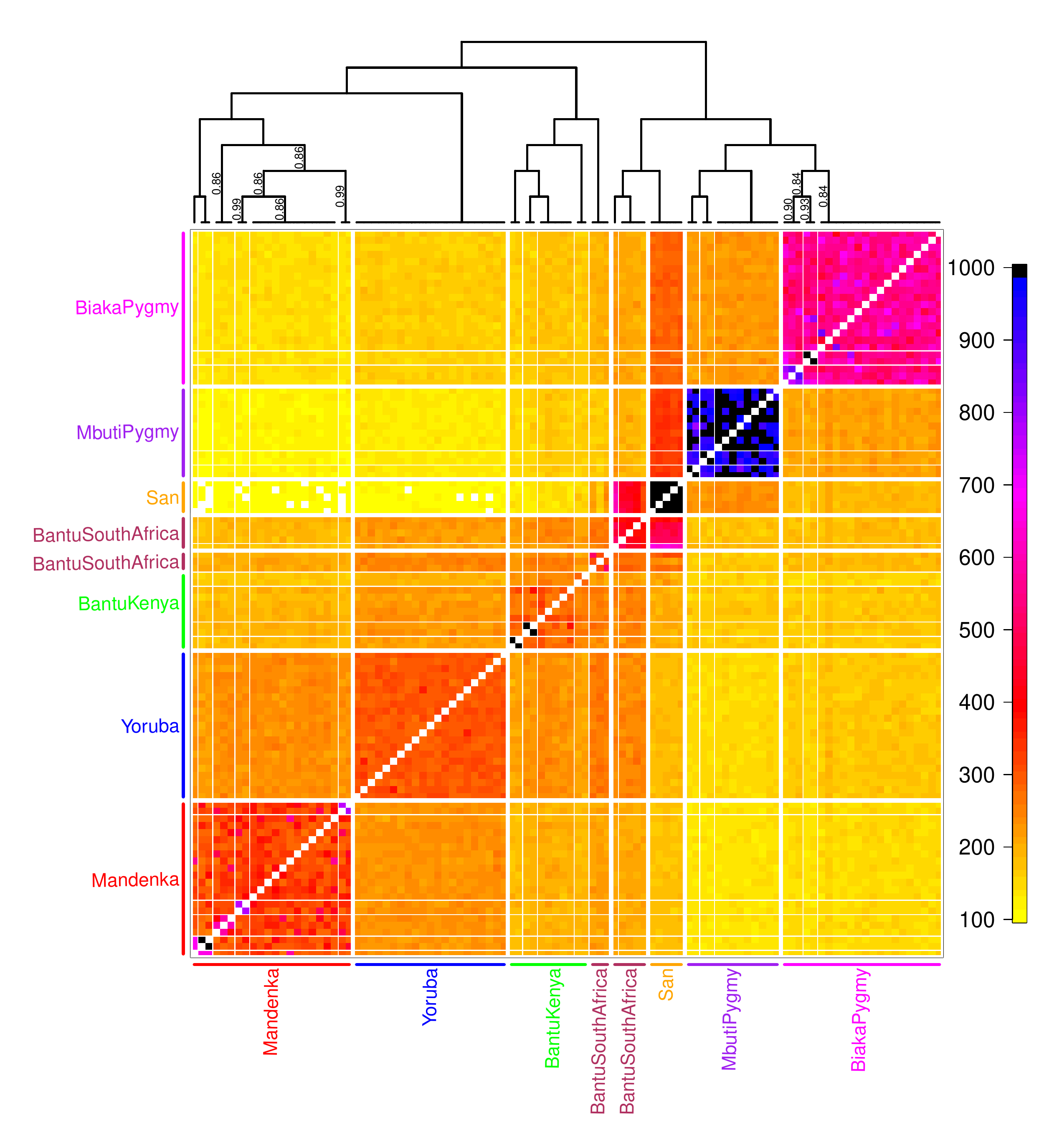}
\vspace{-1cm}
\caption[FineSTRUCTURE Example]{FineSTRUCTURE applied to the African samples of the HGDP dataset*.  The main figure shows the chunk count matrix donated between pairs of individuals, as output by ChromoPainter.  FineSTRUCTURE generates MCMC samples of the population structure, which for clarity are not shown; instead the maximum aposteriori population assignment (i.e. the best single guess) is shown by the white lines which separate populations, and in the top tree: individuals in the same population are in the same lowest level branch. Each edge in the tree produced a bipartition; if this is not perfectly observed in all MCMC samples, the proportion is shown on the tree.  Thick white lines separate major `clades'. A `population' to FineSTRUCTURE is the average of the chunk counts within that population to (columns), and from (rows) all other populations, so within each white square all values should be the same except for noise.  Redrawn from data described in \cite{Lawson2012}.

*\small{\emph{ChromoPainter was applied to the whole HGDP dataset as described in \cite{Lawson2012}, but for simplicity chunks donated to/from other continents are not displayed, although they were used during population assignment.}}
\label{fig:fig4}}
\end{centering}
\end{figure}

Figure \ref{fig:fig4} shows an example of what a population looks like to FineSTRUCTURE, using 102 African individuals from the Human Genetic Diversity Panel (HGDP) - a collection of 1043 individuals from throughout the world \cite{Pickrell09}.  If the `randomly-mating' population model is applicable, all individuals in population $a$ will receive the chunks at the same rate from individuals in population $b$ -- there is no further dependence on who the individual is.  This leads to a `block' structure, clearly visible in the figure, with some labelled populations perfectly described this way (e.g. the Yoruba).  However, FineSTRUCTURE has identified substructure within population labels -- for example, there are related pairs of individuals.  Related individuals are split off in the tree and have high diagonal elements (e.g. within the BantuKenya, the Mandenka, MbutiPygmy and BiakaPygmy).  They can still be identified as being from a particular population because they otherwise share DNA similarly with all of the non-related populations.  Admixture induced substructure can also be seen, e.g. the BantuSouthAfrica consist of two main sub-populations: one related to the BantuKenya, and the other closer to the San (but not the Pygmy populations, who the San are also related to).  In this case the population model is still a useful description, even though it does not directly model the admixture.

The FineSTRUCTURE definition of a population appears very different to that in STRUCTURE - being based on the sharing of haplotypes, rather than the frequency of SNPs.  Yet surprisingly, the model of a population in FineSTRUCTURE applied to unlinked (i.e. ignoring LD, although it may be present) ChromoPainter data is formally equivalent to that in STRUCTURE\footnote{Formally, FineSTRUCTURE and STRUCTURE share the same asymptotic likelihood when SNPs are treated as independent, drift is weak, and number of individuals sampled is large, proof by Simon Myers in the appendix of \cite{Lawson2012}.}.  Despite this strong similarity, there are very important features that differ between the various models available to perform population clustering, which can lead to quite different inferences about populations. The most important is the model for \emph{differences} between populations, which are usually expressed as a model for the population-level parameters. Figure \ref{fig:fig2} shows how these might be important, and Table \ref{tab:popmodels} gives some common approaches.

\begin{table}[h!b!p!]\small
\begin{center}
\caption{Population separation models in popular software\label{tab:popmodels}}
\begin{tabular}{  p{3cm} | p{10cm} }
Software  & Model \\
\hline
STRUCTURE  \cite{Pritchard2000} & SNP frequencies are correlated due to common descent, meaning that each SNP frequency drifts independently from a single ancestral frequency in a `star' structure - each population is independent conditional on the ancestral frequency.  Because the same number of SNP frequencies as SNPs must be inferred, this prior is quite strong. See  \cite{Falush2003} for details.\\
\hline
ADMIXTURE  \cite{Alexander09} & A frequentist method with no explicit population prior, effectively assuming that each population is sampled IID with a uniform SNP distribution. \\
\hline
STRUCTURAMA  \cite{Huelsenbeck2007} & As STRUCTURE, but with a model for the \emph{number} of populations (a Dirichlet Process prior \cite{Ferguson1973}) that is convenient for fast inference.\\
\hline
FineSTRUCTURE  \cite{Lawson2012} & Haplotype sharing rates between different populations are IID with a Multinomial prior.  Because there are only $K$ haplotype rates to infer, this is a weaker prior than used by STRUCTURE.  A Dirichlet Process prior is also used for computational convenience.
\end{tabular}
\end{center}
\end{table}

None of the methods discussed makes powerful use of the difference between individuals being exchangeable within a population, and them being IID.  The closest is the FineSTRUCTURE approach, in which populations are initially inferred on the basis of individuals being drawn IID from a population, but then familial relationships are immediately merged in a `tree-building'\footnote{We omit the details of the tree-building, but broadly, the merging of populations that is `least bad', i.e. causes the lowest increase in the likelihood, is repeatedly applied until only a single population remains.} step, using a heuristic model for kinship.  The heuristic model assumes that related individuals will be placed into the same population and will have additional chunks shared with each other, but the same distribution of other chunks as they would have if the related individual were not in the sample.  Related individuals that share a population can then be merged with their underlying population.  Although this works well in many examples (e.g. Figure \ref{fig:fig4}), the technical details of this two-stage approach are not perfect and new methods are required.

We have described several methods in this section, and claimed that some are `fast' whilst others are not.  However, the speed of these algorithms is strongly dependent on the amount of data available -- `fast' here describes algorithms that scale well with number of genomic loci (and the number of individuals) sampled.
Unfortunately, this is difficult to quantify exactly for MCMC algorithms because how quickly the algorithms converge is dependent on the data.  Broadly though, the ChromoPainter/FineSTRUCTURE approach is fast because the computational cost is linear in the number of loci $L$; the STRUCTURE/STRUCTURAMA approach is at least quadratic.  All these approaches scale similarly with the number of individuals $N$. The exception to these issues is ADMIXTURE which is faster (by roughly a factor of $N$) because it does not perform MCMC.  For reference, the absolute computational cost of running genomic scale datasets (700K SNPs) containing a thousand individuals, is that ChromoPainter/FineSTRUCTURE may take a week to run (although much of this cost is parallelisable); STRUCTURE and STRUCTURAMA would take months, and ADMIXTURE may take only hours/days but usually fails to converge to the global solution, and typically a stable approximation at low $k$ is used. All approaches run accurately within minutes or hours on smaller datasets (say, 10000 SNPs and 100 individuals).

\section{Admixture models}\label{sec:admix}

Both STRUCTURE and ADMIXTURE have a model permitting an individual to relate to multiple populations, known as admixture.  These important models are widely applied in practise as they are typically closer to reality than no-admixture approaches, and -- when LD is ignored -- little more computationally costly. Some high profile examples for Humans are \cite{Rosenberg2002} which examined the human population genetic structure, and \cite{Redon2006,Sladek2007} which analysed functional genetic features, using admixture models to account for the confounding effect of population structure.

\begin{figure}[!ht]
\begin{centering}
\includegraphics[clip,width=1.0\textwidth,angle=0]{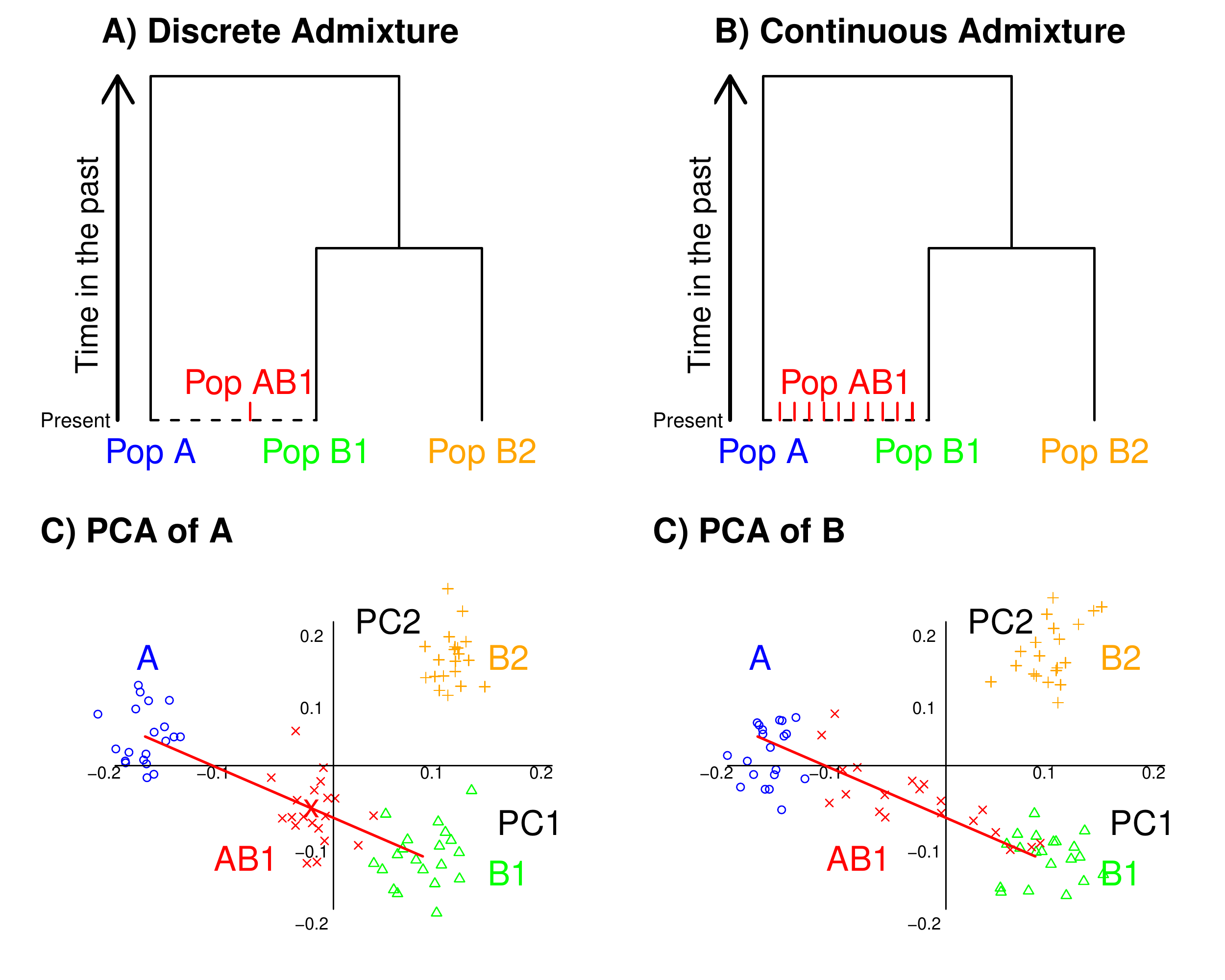}
\caption[Admixture Illustration]{Two simple present-day admixture scenarios for populations split via a tree. A) Discrete admixture, in which two populations are mixed with some characteristic frequency.  Population models can infer an additional population (in this case, AB1) which can either be directly described as a mixture of other populations, or via post-hoc methods such as PCA. B) Continuous admixture, which causes problems for population assignment algorithms. C-D) PC decompositions of the scenarios from A-B.  Admixture fractions will correlate with the position on the red line, shown.  Note however that in both cases, historically admixed populations may have undergone genetic drift.  In this case, there will be additional PC directions in which they are not describable as a mixture.
\label{fig:fig6}}
\end{centering}
\end{figure}

The admixture likelihood of a SNP is the same as in the standard no-admixture model, if we knew which population the SNP came from.  However, an individual is no longer assigned to a single population - each SNP within an individual is viewed as coming from one of $K$ unobserved, `pure' populations, and individuals are seen as mixtures of these populations.  Therefore the model for `what a population is' remains the same, only now we do not observe populations directly, but mixtures from them.  Figure \ref{fig:fig6} illustrates how this can form standard, inferable populations (even without new genetic material in the form of drift) as well as continuous lines in PC space that do not correspond to a population.

Populations are defined similarly to no-admixture models -- as genetically statistically distinct.  However, discrete models will use additional populations to describe mixtures of `real' populations that have undergone genetic drift. As we shall see, if the distribution is complex then admixture models can have multiple interpretations, and a lack of visualisation tools makes assessment more difficult. For example, we cannot plot pure populations in PC space. However, PC analysis can still help visualise the results (Figure \ref{fig:fig6}).

Part of the difficulty in interpreting mixture models for admixture is that they are an incomplete description of how the present day genetic distribution arose.  Differences between pure populations can be described by complex models such as trees \cite{Falush2003}, but the mixing only happens at the present time.  However, admixture may have occurred at any point in history.  For SNP frequency models such as ADMIXTURE and STRUCTURE, the mixture approach captures much of the available information. There are currently no approaches that correctly handle LD and admixture in general.  A problem is that the length of LD blocks contains strong information about admixture dates, which would ideally be used in an admixture description.  Therefore current admixture models need completely revising to utilise this information.

To construct admixture models from LD data, it becomes necessary to form specific hypotheses about historical drift and admixture scenarios.  These include, how do populations separate and merge? How is DNA shared between individuals in different populations via migration?  Such a model might resemble the generative models described in Section \ref{sec:gen}, or might take inferred populations from STRUCTURE/FineSTRUCTURE and explain their relationship \emph{post-hoc}. TreeMix \cite{Pickrell2012} is one method for this; it describes the relationship between populations as a tree, subject to small amounts of historical admixture.  However, post-hoc approaches typically lose information, both by discarding information present in the raw data, and by discarding uncertainty present in the first stage of modelling.  Conversely, two-stage approaches are usually faster, and in some cases (as discussed for ChromoPainter/FineSTRUCTURE) it can be shown that no information is lost.

\section{Population-free approaches} \label{sec:popfree}

We may view a `population' as a somewhat arbitrary boundary between local variation and larger scale differences of relevance to real world questions.  If we are willing to abandon neat segregation of individuals into populations, we can provide a more accurate description of the underlying genetic structure, at the price of interpretability.  This trades off one purpose of population modelling -- aiding understanding of the underlying variation -- with accurately representing the underlying genetic structure in a low-dimensional representation.

STRUCTURE-like models are `semi-parametric' because they use a parametric description of a population (e.g. a SNP frequency or a haplotype sharing distribution based on results from a generative model of how variation arises) but allow an arbitrary number of populations to describe the sample.  However, they penalise the use of too many populations in order to obtain interpretable results, and the description of a population is fully parametric.  Non-parametric methods\footnote{`Non-parametric' methods describe data using an infinite-dimensional representation in an appropriately chosen basis; of course only a finite number of components are used in practise, which is typically equal to or less than the number of data points.} (e.g. \cite{Wasserman2005}) can provide a more accurate representation of the underlying genetic structure than parametric methods when modelling assumptions are invalid.  For example, real populations are often continuously admixed over many timescales, and have complex substructure.  This typically leads to methods such as FineSTRUCTURE inferring a large number of populations that are hard to interpret.  We could consider a non-parametric description of the populations themselves, and potentially obtain simpler results.

\begin{figure}[!ht]
\begin{centering}
\includegraphics[clip,width=1.0\textwidth,angle=0]{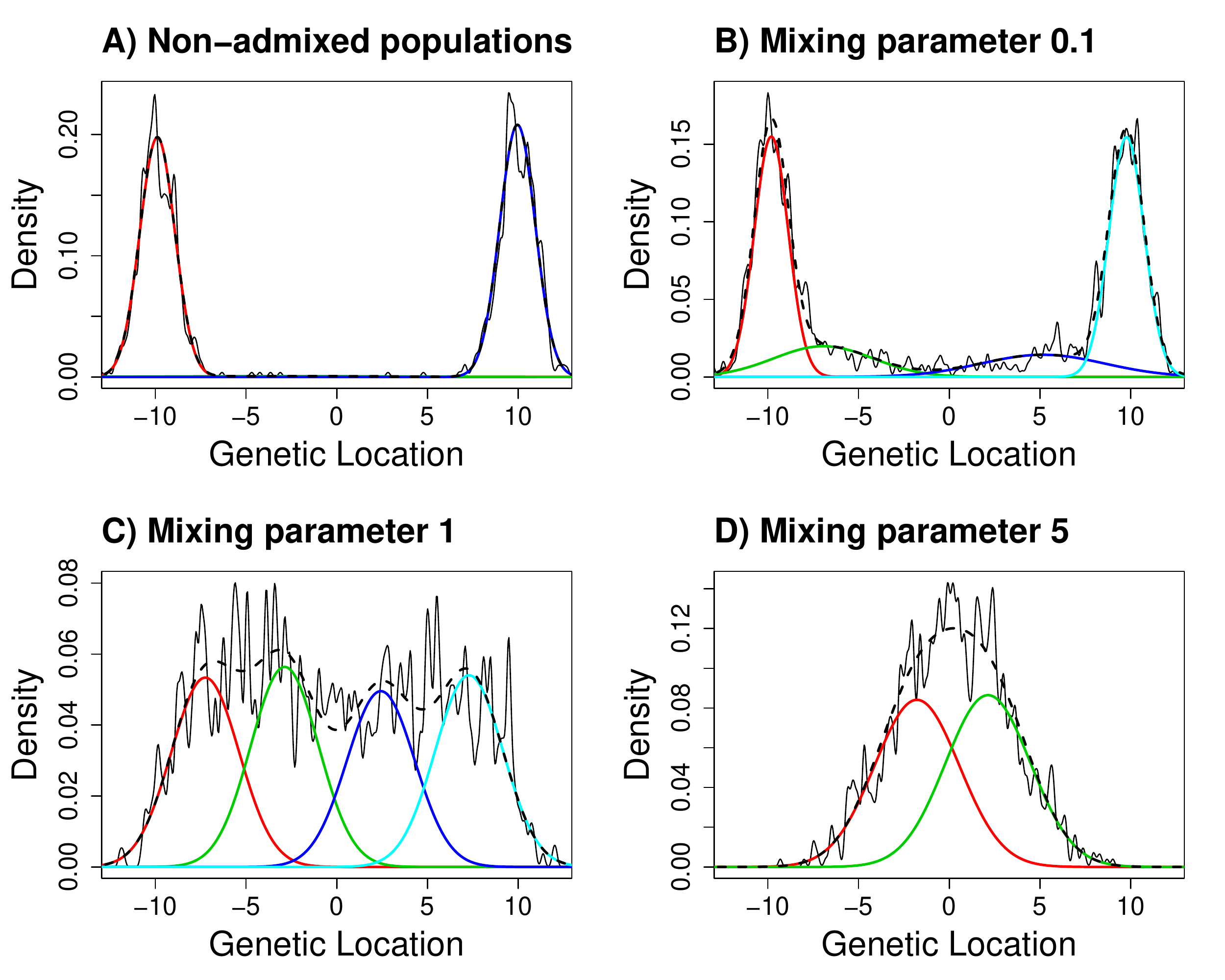}
\caption[Admixture Illustration 2]{The effect of admixture on population inference in `genetic space' (equivalent to the first PC of a PC analysis of these two populations).  Mixtures between two randomly mating populations were generated (solid wiggly line) by sampling individual's proportions from a $\mathrm{Dirichlet}(p,p)$ distribution, for A) $p=0.01$, B) $p=0.1$, C) $p=1$, and D) $p=5$.  Low values of $p$ mean that individuals predominantly are from a single population; high values mean they are close to equal proportions of both.  Also shown is the `best' no-admixture population assignment*
(`bell' curves) and the resulting density this generates (black dashed line).  Note that population methods infer `spurious' populations to explain the intermediate types (B-C).  In D), there is a `single' population but it has complex structure and is inferred as two populations.  This is inferable because although there is a single mode, the distribution is not Normal and the variance is inflated. 

*\small{\emph{Population assignment performed using MCLUST \cite{RafteryMclust} as described in \cite{LawsonFalush2012}. 1000 individuals were sampled; `Populations' here are Normal distributions representing a hypothetical first PC (e.g. \cite{McVean09}).}}
\label{fig:fig7}}
\end{centering}
\end{figure}

An interesting feature of admixture models is that they can correct for inadequacies in the conceptual model a population.  By allowing a population to be described as a mixture of (very) many unobserved components, any genetic distribution can be represented. This is essentially a non-parametric statistical view of a population.  However, this view is not commonly used for two reasons.  Firstly, it confuses the theoretical population concept with the observed populations, so that it is no longer clear what a `population' is.  Secondly, a lot of data is required to infer a large number of mixture components\footnote{The number of samples is exponential in the number of mixture components, when they overlap. This makes inferring more than a few components theoretically `impossible'.} \cite{Kalai2012}.  Inference algorithms frequently get trapped at suboptimal population mixtures, leading to repeated runs of the same algorithm finding different solutions.  Figure \ref{fig:fig7} illustrates how admixture appears in inference.  Admixture components are not always associable with either a cluster nor a peak in the density in a PCA decomposition \cite{Carreira-Perpinan2003} (Figure \ref{fig:fig7}D).  The drift and admixture fractions in this heavily admixed scenario are very hard to separate and lead to unstable estimates. The end result is that admixture models robustly identify only a small number of populations which may not give a good `non-parametric' description of the data and can still be hard to interpret.  If multiple pure populations are mixed, this becomes even more complex.

Species occupying continuous habitat ranges may require a spatially explicit model.
\cite{Guillot2009} reviews isolation-by-distance models, which allow for individuals to be distributed at any point over a spatial region. `Random fields' can be used to describe how populations change over space, either using discrete populations \cite{Guillot2005,Francois2006,Corander2008}, or admixture \cite{Durand2009}. This approach allows for very flexible specification of how spatial correlations arise (a popular choice is the Markov random field \cite{Kindermann1980}).  These models resemble the approaches described above but use geographical distance to help explain differences between populations.  Admixture models are more interpretable when using spatial information, and may be more robust \cite{FrancoisDurand2010}, as additional information is available to infer mixture components. However, general application requires care in case spatial correlations are not the primary source of variation.  While inference in this framework is very challenging \cite{Guillot2009a} there has been significant recent progress in the technical methods \cite{Grelaud2009,Everitt2012}.

Although this section has taken a negative view of the population concept, in practise many species can be correctly modelled with it.  Fragmented habitats due to man-made or geographical barriers to migration are naturally represented by discrete populations, and spatial structure can be handled either via standard methods or networks of demes \cite{Fortuna2009}.  Historical populations are only accessible via rare fossil DNA and are therefore rarely continuous in practise. Population models will therefore coexist with approaches that define genetic distributions implicitly and both may be suitable for different questions even for the same species.

\section{Discussion}

There are many features on which we may wish to group individuals into `populations' - language, culture, geography and history will all lead to different groupings suitable for different purposes. Here we focus on population genetic approaches that group all individuals which cannot be statistically distinguished into the same population.  This is an appealing approach since it is not necessary to define how population differences arise - we only need a model for what it means to be in a population. A natural model for a population is a randomly mating group of individuals, which allows for powerful theoretical results to be applied.

There are two main reasons to search for population structure in the genetics of individuals.  The first is for understanding; we wish to classify the world around us in the simplest way possible (but no simpler).  This is the main motivation behind the generative models from Section \ref{sec:gen}, and the statistical approaches from Section \ref{sec:stat} that tie into these.  We describe present day individuals using populations so that we can then reason about the historical forces of separation, migration, admixture, etc. that have influenced the populations we observe. However, the second major application is to remove population structure as a confounder of genetic correlation, for the understanding of genomic function.  Because genomic variation occurs at far more loci than the number of individuals we can hope to sample, naive inference is confounded by many factors including population structure.  In this case, the nonparametric viewpoint (Section \ref{sec:popfree}) is appropriate.  Major population structure approaches \cite{Tiwari2008} typically use either PC analysis \cite{Price2006} or STRUCTURE-like models for this.  Neither fully captures complex population structure, particularly when provided with linked data.  Although linkage mapping approaches exist (e.g. \cite{Redden2006}), these can only account for simple population structure and we expect to see new methods for detecting genetic selection based on linked data.

We have reviewed the main population assignment approaches for weakly linked genetic data (e.g. STRUCTURE \cite{Pritchard2000}) and seen that populations in such models have a clear interpretation in terms of generative models for how genetic variation arises within a population.  Models of differences between populations can be used where appropriate to gain further statistical power, and then the model is fully generative.

When data are densely sampled from the genome, handling Linkage Disequilibrium becomes a necessity.  For population identification, our ChromoPainter and FineSTRUCTURE \cite{Lawson2012} approach is the most powerful currently available.  Here, individuals are first summarised in terms of the number of contiguous haplotype blocks they share with each other, after which they can be clustered. This approach theoretically matches previous SNP models but has dramatically improved computational performance, as well as providing a more in depth description of the data.

Unfortunately, modern methods for population inference have \emph{too much} power to infer randomly mating populations, and can potentially find as many populations as there are individuals.  To address this, we have discussed some relaxations from random mating which can be applied when defining a population.  Some unambiguous `within-population' variation is familial structure, which is well understood in principle, and may fit within standard statistical methodology.  A more difficult extension is admixture between sub-populations, which is difficult because there is no line between a `sub-population' and a `true population'.  Although such deviations may be hard to either practically or theoretically distinguish from differences between populations, we remain positive about the utility for theoretical population concepts in understanding genetic variation. 

\bibliographystyle{mychicago}
\bibliography{populationdefinition}

\begin{thebibliography}{}

\bibitem[\protect\citeauthoryear{Alexander, Novembre, and
  Lange}{\textsc{Alexander} {\em et~al.\@}}{2009}]{Alexander09}
\textsc{Alexander, D.~H.}, \textsc{J.~Novembre}, and \textsc{K.~Lange}, 2009\ \
  Fast model-based estimation of ancestry in unrelated individuals.
\newblock Genome Research~{\bf 19}\textbf{:}\ 1655--1664.

\bibitem[\protect\citeauthoryear{Ammerman and Cavalli-Sforza}{\textsc{Ammerman}
  and \textsc{Cavalli-Sforza}}{1984}]{AmmermanCavalli-Sforza1984}
\textsc{Ammerman, A.} and \textsc{L.~L. Cavalli-Sforza}, 1984\ \ {\em The
  Neolithic Transition and the Genetics of Populations in Europe}.
\newblock Princeton: Princeton Univ. Press.

\bibitem[\protect\citeauthoryear{Beaumont}{\textsc{Beaumont}}{2010}]{Beaumont2%
010}
\textsc{Beaumont, M.}, 2010\ \ Approximate Bayesian Computation in Evolution
  and Ecology.
\newblock Annual Review of Ecology, Evolution, and Systematics~{\bf
  41}\textbf{:}\ 379--406.

\bibitem[\protect\citeauthoryear{Bernardo}{\textsc{Bernardo}}{1996}]{Bernardo1%
996}
\textsc{Bernardo, J.~M.}, 1996\ \ The Concept of Exchangeability and its
  Applications.
\newblock Far East Journal of Mathematical Sciences~{\bf 4}\textbf{:}\ 111–121.

\bibitem[\protect\citeauthoryear{Bonin, Taberlet, Miaud, and
  Pompanon}{\textsc{Bonin} {\em et~al.\@}}{2006}]{Bonin2006}
\textsc{Bonin, A.}, \textsc{P.~Taberlet}, \textsc{C.~Miaud}, and
  \textsc{F.~Pompanon}, April 2006\ \ Explorative Genome Scan to Detect
  Candidate Loci for Adaptation Along a Gradient of Altitude in the Common Frog
  (Rana temporaria).
\newblock Molecular Biology and Evolution~{\em 23\/}(4)\textbf{:}\ 773--783.

\bibitem[\protect\citeauthoryear{Carreira-Perpinan and
  Williams}{\textsc{Carreira-Perpinan} and
  \textsc{Williams}}{2003}]{Carreira-Perpinan2003}
\textsc{Carreira-Perpinan, M.} and \textsc{C.~Williams}, 2003\ \ On the number
  of modes of a Gaussian mixture.
\newblock In {\em Scale Space Methods in Computer Vision}, pp.\  625--640.

\bibitem[\protect\citeauthoryear{Chikhi, Nichols, Barbujani, and
  Beaumont}{\textsc{Chikhi} {\em et~al.\@}}{2002}]{Chikhi2002}
\textsc{Chikhi, L.}, \textsc{R.~A. Nichols}, \textsc{G.~Barbujani}, and
  \textsc{M.~A. Beaumont}, 2002\ \ Y genetic data support the Neolithic demic
  diffusion model.
\newblock {PNAS}~{\bf 99}\textbf{:}\ 11008--11013.

\bibitem[\protect\citeauthoryear{Corander, Sir\'{e}n, and
  Arjas}{\textsc{Corander} {\em et~al.\@}}{2008}]{Corander2008}
\textsc{Corander, J.}, \textsc{J.~Sir\'{e}n}, and \textsc{E.~Arjas}, 2008\ \
  Bayesian spatial modeling of genetic population structure.
\newblock Computational Statistics~{\bf 23}\textbf{:}\ 111--129.

\bibitem[\protect\citeauthoryear{Cornuet, Santos, Beaumont, Robert, Marin,
  Balding, Guillemaud, and Estoup}{\textsc{Cornuet} {\em
  et~al.\@}}{2008}]{Cornuet2008}
\textsc{Cornuet, J.-M.}, \textsc{F.~Santos}, \textsc{M.~A. Beaumont},
  \textsc{C.~P. Robert}, \textsc{J.-M. Marin}, \textsc{D.~J. Balding},
  \textsc{T.~Guillemaud}, and \textsc{A.~Estoup}, 2008\ \ Inferring population
  history with DIY ABC: a user-friendly approach to approximate Bayesian
  computation.
\newblock Bioinformatics~{\em 24\/}(23)\textbf{:}\ 2713--2719.

\bibitem[\protect\citeauthoryear{Creighton and McClintock}{\textsc{Creighton}
  and \textsc{McClintock}}{1931}]{CreightonMcClintock1931}
\textsc{Creighton, H.} and \textsc{B.~McClintock}, 1931\ \ A Correlation of
  Cytological and Genetical Crossing-Over in Zea Mays.
\newblock Proc. Natl. Acad. Sci. USA~{\bf 17}\textbf{:}\ 492--497.

\bibitem[\protect\citeauthoryear{Dawson and Belkhir}{\textsc{Dawson} and
  \textsc{Belkhir}}{2001}]{Dawson2001}
\textsc{Dawson, K.~J.} and \textsc{K.~Belkhir}, 2001\ \ A Bayesian approach to
  the identification of panmictic populations and the assignment of
  individuals.
\newblock Genetical Research~{\bf 78}\textbf{:}\ 59--77.

\bibitem[\protect\citeauthoryear{Durand, Jay, Gaggiotti, and
  François}{\textsc{Durand} {\em et~al.\@}}{2009}]{Durand2009}
\textsc{Durand, E.}, \textsc{F.~Jay}, \textsc{O.~Gaggiotti}, and
  \textsc{O.~François}, 2009\ \ Spatial inference of admixture proportions and
  secondary contact zones.
\newblock Molecular Biology and Evolution~{\bf 26}\textbf{:}\ 1963--197.

\bibitem[\protect\citeauthoryear{Durand, Patterson, Reich, and
  Slatkin}{\textsc{Durand} {\em et~al.\@}}{2011}]{Durand2011}
\textsc{Durand, E.~Y.}, \textsc{N.~Patterson}, \textsc{D.~Reich}, and
  \textsc{M.~Slatkin}, 2011\ \ Testing for Ancient Admixture between Closely
  Related Populations.
\newblock Mol. Biol. Evol.~{\bf 28}\textbf{:}\ 2239--2252.

\bibitem[\protect\citeauthoryear{Everitt}{\textsc{Everitt}}{2012}]{Everitt2012}
\textsc{Everitt, R.}, 2012\ \ Bayesian Parameter Estimation for Latent Markov
  Random Fields and Social Networks.
\newblock to appear in Journal of Computational and Graphical Statistics.

\bibitem[\protect\citeauthoryear{Ewing and Hermisson}{\textsc{Ewing} and
  \textsc{Hermisson}}{2010}]{EwingHermisson2010}
\textsc{Ewing, G.} and \textsc{J.~Hermisson}, 2010\ \ {MSMS}: a coalescent
  simulation program including recombination, demographic structure and
  selection at a single locus.
\newblock Bioinformatics~{\em 26\/}(16)\textbf{:}\ 2064--2065.

\bibitem[\protect\citeauthoryear{Falush, Stephens, and
  Pritchard}{\textsc{Falush} {\em et~al.\@}}{2003}]{Falush2003}
\textsc{Falush, D.}, \textsc{M.~Stephens}, and \textsc{J.~K. Pritchard}, 2003\
  \ Inference of Population Structure Using Multilocus Genotype Data: Linked
  Loci and Correlated Allele Frequencies.
\newblock Genetics~{\bf 164}\textbf{:}\ 1567--1587.

\bibitem[\protect\citeauthoryear{Ferguson}{\textsc{Ferguson}}{1973}]{Ferguson1%
973}
\textsc{Ferguson, T.~S.}, 1973\ \ A Bayesian analysis of some nonparametric
  problems.
\newblock Ann. Statist.~{\bf 1}\textbf{:}\ 209--230.

\bibitem[\protect\citeauthoryear{Fortuna, Albaladejo, Fern\'{a}ndez, Aparicio,
  and Bascompte}{\textsc{Fortuna} {\em et~al.\@}}{2009}]{Fortuna2009}
\textsc{Fortuna, M.~A.}, \textsc{R.~G. Albaladejo}, \textsc{L.~Fern\'{a}ndez},
  \textsc{A.~Aparicio}, and \textsc{J.~Bascompte}, 2009\ \ Networks of spatial
  genetic variation across species.
\newblock PNAS~{\bf 106}\textbf{:}\ 19044–19049.

\bibitem[\protect\citeauthoryear{Fraley and Raftery}{\textsc{Fraley} and
  \textsc{Raftery}}{2009}]{RafteryMclust}
\textsc{Fraley, C.} and \textsc{A.~E. Raftery}, 2006 (revised 2009)\ \ MCLUST
  Version 3 for R: Normal Mixture Modeling and Model-Based Clustering.
\newblock Technical Report 504, University of Washington.

\bibitem[\protect\citeauthoryear{Fran\c{c}ois, Ancelet, and
  Guillot}{\textsc{Fran\c{c}ois} {\em et~al.\@}}{2006}]{Francois2006}
\textsc{Fran\c{c}ois, O.}, \textsc{S.~Ancelet}, and \textsc{G.~Guillot}, 2006\
  \ Bayesian Clustering Using Hidden Markov Random Fields in Spatial Population
  Genetics.
\newblock Genetics~{\bf 174}\textbf{:}\ 805--816.

\bibitem[\protect\citeauthoryear{Fran\c{c}ois, Currat, Ray, Han, Excoffier, and
  Novembre}{\textsc{Fran\c{c}ois} {\em et~al.\@}}{2010}]{Francois2010}
\textsc{Fran\c{c}ois, O.}, \textsc{M.~Currat}, \textsc{N.~Ray},
  \textsc{E.~Han}, \textsc{L.~Excoffier}, and \textsc{J.~Novembre}, 2010\ \
  Principal Component Analysis under Population Genetic Models of Range
  Expansion and Admixture.
\newblock Mol Biol Evol~{\bf 27}\textbf{:}\ 1257--1268.

\bibitem[\protect\citeauthoryear{Fran\c{c}ois and Durand}{\textsc{Fran\c{c}ois}
  and \textsc{Durand}}{2010}]{FrancoisDurand2010}
\textsc{Fran\c{c}ois, O.} and \textsc{E.~Durand}, 2010\ \ Spatially explicit
  Bayesian clustering models in population genetics.
\newblock Molecular Ecology Resources~{\bf 10}\textbf{:}\ 773--784.

\bibitem[\protect\citeauthoryear{Gamerman}{\textsc{Gamerman}}{1997}]{Gamerman1%
997}
\textsc{Gamerman, D.}, 1997\ \ {\em Markov Chain Monte Carlo: Stochastic
  simulation for Bayesian inference}.
\newblock 2-6 Boundary Row, London, UK. SE1 8HN: Chapman and Hall.

\bibitem[\protect\citeauthoryear{Grelaud, Robert, and Marin}{\textsc{Grelaud}
  {\em et~al.\@}}{2009}]{Grelaud2009}
\textsc{Grelaud, A.}, \textsc{C.~P. Robert}, and \textsc{J.-M. Marin}, 2009\ \
  ABC methods for model choice in Gibbs random fields.
\newblock C. R. Acad. Sci. Paris~{\bf 347}\textbf{:}\ 205--210.

\bibitem[\protect\citeauthoryear{Griffiths}{\textsc{Griffiths}}{1981}]{Griffit%
hs1981}
\textsc{Griffiths, R.~C.}, 1981\ \ Neutral two-locus multiple allele models
  with recombination.
\newblock Theor. Popul. Biol.~{\bf 19}\textbf{:}\ 169--186.

\bibitem[\protect\citeauthoryear{Guillot}{\textsc{Guillot}}{2009}]{Guillot2009%
a}
\textsc{Guillot, G.}, 2009\ \ On the inference of spatial structure from
  population genetics data.
\newblock Bioinformatics~{\bf 25}\textbf{:}\ 1796--1801.

\bibitem[\protect\citeauthoryear{Guillot, Estoup, and Mortier}{\textsc{Guillot}
  {\em et~al.\@}}{2005}]{Guillot2005}
\textsc{Guillot, G.}, \textsc{A.~Estoup}, and \textsc{F.~Mortier}, 2005\ \ A
  spatial statistical model for landscape genetics.
\newblock Genetics~{\bf 170}\textbf{:}\ 1261–--1280.

\bibitem[\protect\citeauthoryear{Guillot, Leblois, Coulon, and
  Frantz}{\textsc{Guillot} {\em et~al.\@}}{2009}]{Guillot2009}
\textsc{Guillot, G.}, \textsc{R.~Leblois}, \textsc{A.~Coulon}, and
  \textsc{A.~Frantz}, 2009\ \ Statistical methods in spatial genetics.
\newblock Molecular Ecology~{\bf 18}\textbf{:}\ 4734–--4756.

\bibitem[\protect\citeauthoryear{Hein, Schierup, and Wiuf}{\textsc{Hein} {\em
  et~al.\@}}{2005}]{Hein2005}
\textsc{Hein, J.}, \textsc{M.~H. Schierup}, and \textsc{C.~Wiuf}, 2005\ \ {\em
  Gene Genealogies, Variation and Evolution: A Primer in Coalescent Theory}.
\newblock Oxford: Oxford University Press.

\bibitem[\protect\citeauthoryear{Henderson and Lambert}{\textsc{Henderson} and
  \textsc{Lambert}}{1982}]{HendersonLambert1982}
\textsc{Henderson, N.~R.} and \textsc{D.~M. Lambert}, 1982\ \ No significant
  deviation from random mating of worldwide populations of Drosophila
  melanogaster.
\newblock Nature~{\bf 300}\textbf{:}\ 437--440.

\bibitem[\protect\citeauthoryear{Herrera and Bazaga}{\textsc{Herrera} and
  \textsc{Bazaga}}{2008}]{HerreraBazaga2008}
\textsc{Herrera, C.~M.} and \textsc{P.~Bazaga}, 2008\ \ Population-genomic
  approach reveals adaptive floral divergence in discrete populations of a hawk
  moth-pollinated violet.
\newblock Mol Ecol.~{\bf 17}\textbf{:}\ 5378--5390.

\bibitem[\protect\citeauthoryear{Hudson}{\textsc{Hudson}}{1983}]{Hudson1983}
\textsc{Hudson, R.~R.}, 1983\ \ Properties of the neutral model with intragenic
  recombination.
\newblock Theor.Pop.Biol.~{\bf 23}\textbf{:}\ 213--201.

\bibitem[\protect\citeauthoryear{Huelsenbeck and
  Andolfatto}{\textsc{Huelsenbeck} and
  \textsc{Andolfatto}}{2007}]{Huelsenbeck2007}
\textsc{Huelsenbeck, J.~P.} and \textsc{P.~Andolfatto}, 2007\ \ Inference of
  Population Structure Under a Dirichlet Process Model.
\newblock Genetics~{\bf 175}\textbf{:}\ 1787--1802.

\bibitem[\protect\citeauthoryear{{International~HapMap~Consortium}}{\textsc{{I%
nternational~HapMap~Consortium}}}{2007}]{Hapmap2007}
\textsc{{International~HapMap~Consortium}}, 2007\ \ A second generation human
  haplotype map of over 3.1 million SNPs.
\newblock Nature~{\bf 449}\textbf{:}\ 851--861.

\bibitem[\protect\citeauthoryear{Johnson and Kotz}{\textsc{Johnson} and
  \textsc{Kotz}}{1977}]{JohnsonKotz1977}
\textsc{Johnson, N.} and \textsc{S.~Kotz}, 1977\ \ {\em Urn Models and Their
  Application}.
\newblock John Wiley.

\bibitem[\protect\citeauthoryear{Kadanoff}{\textsc{Kadanoff}}{2000}]{KadanoffS%
tatPhys2000}
\textsc{Kadanoff, L.~P.}, 2000\ \ {\em Statistical Physics: Statics, Dynamics
  and Renormalization}.
\newblock World Scientific.

\bibitem[\protect\citeauthoryear{Kalai, Moitra, and Valiant}{\textsc{Kalai}
  {\em et~al.\@}}{2012}]{Kalai2012}
\textsc{Kalai, A.~T.}, \textsc{A.~Moitra}, and \textsc{G.~Valiant}, 2012\ \
  Disentangling Gaussians.
\newblock In {\em Communications of the ACM}, pp.\  113--120.

\bibitem[\protect\citeauthoryear{Kindermann and Snell}{\textsc{Kindermann} and
  \textsc{Snell}}{1980}]{Kindermann1980}
\textsc{Kindermann, R.} and \textsc{J.~L. Snell}, 1980\ \ {\em Markov Random
  Fields and Their Applications}.
\newblock American Mathematical Society.

\bibitem[\protect\citeauthoryear{Kingman}{\textsc{Kingman}}{1982}]{Kingman1982}
\textsc{Kingman, J. F.~C.}, 1982\ \ {The coalescent}.
\newblock Stochastic Processes and their Applications~{\bf 13}\textbf{:}\
  235--248.

\bibitem[\protect\citeauthoryear{Knowles}{\textsc{Knowles}}{2009}]{Knowles2009}
\textsc{Knowles, L.~L.}, 2009\ \ Statistical Phylogeography.
\newblock Annual Review of Ecology, Evolution, and Systematics~{\bf
  40}\textbf{:}\ 593--612.

\bibitem[\protect\citeauthoryear{Knowles and Maddison}{\textsc{Knowles} and
  \textsc{Maddison}}{2002}]{KnowlesMaddison2002}
\textsc{Knowles, L.~L.} and \textsc{W.~P. Maddison}, 2002\ \ Statistical
  phylogeography.
\newblock Mol. Ecol.~{\bf 11}\textbf{:}\ 2623--2635.

\bibitem[\protect\citeauthoryear{Kyriazopoulou-Panagiotopoulou, Haghighi,
  Aerni, Sundquist, Bercovici, and
  Batzoglou}{\textsc{Kyriazopoulou-Panagiotopoulou} {\em
  et~al.\@}}{2011}]{Kyria2011}
\textsc{Kyriazopoulou-Panagiotopoulou, S.}, \textsc{D.~K. Haghighi},
  \textsc{S.~J. Aerni}, \textsc{A.~Sundquist}, \textsc{S.~Bercovici}, and
  \textsc{S.~Batzoglou}, 2011\ \ Reconstruction of genealogical relationships
  with applications to Phase III of HapMap.
\newblock Bioinformatics~{\bf 27}\textbf{:}\ i333--i341.

\bibitem[\protect\citeauthoryear{Lawson and Falush}{\textsc{Lawson} and
  \textsc{Falush}}{2012}]{LawsonFalush2012}
\textsc{Lawson, D.~J.} and \textsc{D.~Falush}, 2012\ \ Population
  Identification Using Genetic Data.
\newblock To appear Ann. Rev. Genom. and Hum. Genet., available online before
  publication at \url{http://www.annualreviews.org}.

\bibitem[\protect\citeauthoryear{Lawson, Hellenthal, Myers, and
  Falush}{\textsc{Lawson} {\em et~al.\@}}{2012}]{Lawson2012}
\textsc{Lawson, D.~J.}, \textsc{G.~Hellenthal}, \textsc{S.~Myers}, and
  \textsc{D.~Falush}, 2012\ \ Inference of population structure using dense
  haplotype data.
\newblock {PLoS Genetics}~{\bf 8}\textbf{:}\ e100245.

\bibitem[\protect\citeauthoryear{Lawson and Jensen}{\textsc{Lawson} and
  \textsc{Jensen}}{2007}]{LawsonJensen2007}
\textsc{Lawson, D.~J.} and \textsc{H.~J. Jensen}, 2007\ \ Neutral Evolution in
  a Biological Population as Diffusion in Phenotype Space: Reproduction with
  Local Mutation but without Selection.
\newblock Phys. Rev. Lett.~{\bf 98}\textbf{:}\ 098102.

\bibitem[\protect\citeauthoryear{Li and Stephens}{\textsc{Li} and
  \textsc{Stephens}}{2003}]{LiStephens2003}
\textsc{Li, N.} and \textsc{M.~Stephens}, 2003\ \ Modeling Linkage
  Disequilibrium and Identifying Recombination Hotspots Using Single-Nucleotide
  Polymorphism Data.
\newblock Genetics~{\bf 165}\textbf{:}\ 2213--2233.

\bibitem[\protect\citeauthoryear{Lifshitz}{\textsc{Lifshitz}}{1952}]{Lifshitz1%
952}
\textsc{Lifshitz, I.~M.}, 1952\ \ On a problem of the theory of perturbations
  connected with quantum statistics.
\newblock Uspekhi Mat. Nauk~{\bf 7}\textbf{:}\ 171--180.

\bibitem[\protect\citeauthoryear{Lynch}{\textsc{Lynch}}{2010}]{Lynch2010}
\textsc{Lynch, M.}, 2010\ \ Evolution of the mutation rate.
\newblock Trends in Genetics~{\bf 26}\textbf{:}\ 345--–352.

\bibitem[\protect\citeauthoryear{Mardis}{\textsc{Mardis}}{2011}]{Mardis2011}
\textsc{Mardis, E.}, 2011\ \ A decade's perspective on DNA sequencing
  technology.
\newblock Nature~{\bf 470}\textbf{:}\ 198--203.

\bibitem[\protect\citeauthoryear{McVean}{\textsc{McVean}}{2009}]{McVean09}
\textsc{McVean, G.}, 2009\ \ {A Genealogical Interpretation of Principal
  Components Analysis}.
\newblock {PLoS Genetics}~{\em 5\/}(10)\textbf{:}\ e1000686.

\bibitem[\protect\citeauthoryear{Milgroom}{\textsc{Milgroom}}{1996}]{Milgroom1%
996}
\textsc{Milgroom, M.~G.}, 1996\ \ Recombination and the multilocus structure of
  fungal populations.
\newblock Annual Review of Phytopathology~{\bf 34}\textbf{:}\ 457--477.

\bibitem[\protect\citeauthoryear{Myers, Bowden, Tumian, Bontrop, Freeman,
  MacFie, McVean, and Donnelly}{\textsc{Myers} {\em
  et~al.\@}}{2010}]{Myers2010}
\textsc{Myers, S.}, \textsc{R.~Bowden}, \textsc{A.~Tumian}, \textsc{R.~E.
  Bontrop}, \textsc{C.~Freeman}, \textsc{T.~S. MacFie}, \textsc{G.~McVean}, and
  \textsc{P.~Donnelly}, 2010\ \ Drive Against Hotspot Motifs in Primates
  Implicates the PRDM9 Gene in Meiotic Recombination.
\newblock Science~{\bf 327}\textbf{:}\ 876--879.

\bibitem[\protect\citeauthoryear{Pemberton, Wang, Li, and
  Rosenberg}{\textsc{Pemberton} {\em et~al.\@}}{2010}]{Pemberton2010}
\textsc{Pemberton, T.~J.}, \textsc{C.~Wang}, \textsc{J.~Z. Li}, and
  \textsc{N.~A. Rosenberg}, 2010\ \ Inference of Unexpected Genetic Relatedness
  among Individuals in HapMap Phase III.
\newblock Am. J. Hum. Genet.~{\bf 87}\textbf{:}\ 457--464.

\bibitem[\protect\citeauthoryear{Pickrell, Coop, Novembre, Kudaravalli, Li,
  Absher, Srinivasan, Barsh, Myers, Feldman, and Pritchard}{\textsc{Pickrell}
  {\em et~al.\@}}{2009}]{Pickrell09}
\textsc{Pickrell, J.~K.}, \textsc{G.~Coop}, \textsc{J.~Novembre},
  \textsc{S.~Kudaravalli}, \textsc{J.~Z. Li}, \textsc{D.~Absher}, \textsc{B.~S.
  Srinivasan}, \textsc{G.~S. Barsh}, \textsc{R.~M. Myers}, \textsc{M.~W.
  Feldman}, and \textsc{J.~K. Pritchard}, 2009\ \ Signals of recent positive
  selection in a worldwide sample of human populations.
\newblock Genome Research~{\bf 19}\textbf{:}\ 826--37.

\bibitem[\protect\citeauthoryear{Pickrell and Pritchard}{\textsc{Pickrell} and
  \textsc{Pritchard}}{2012}]{Pickrell2012}
\textsc{Pickrell, J.~K.} and \textsc{J.~K. Pritchard}, 2012\ \ Inference of
  population splits and mixtures from genome-wide allele frequency data.
\newblock arXiv:1206.2332v1.

\bibitem[\protect\citeauthoryear{Platt, Horton, Huang, Li, Anastasio, Mulyati,
  \r{A}gren, Bossdorf, Byers, Kathleen, Dunning, Holub, Hudson, Corre, Loudet,
  Roux, Warthmann, Weigel, Rivero, Scholl, Nordborg, Bergelson, and
  Borevitz}{\textsc{Platt} {\em et~al.\@}}{2010}]{Platt2010}
\textsc{Platt, A.}, \textsc{M.~Horton}, \textsc{Y.~S. Huang}, \textsc{Y.~Li},
  \textsc{A.~E. Anastasio}, \textsc{N.~W. Mulyati}, \textsc{J.~\r{A}gren},
  \textsc{O.~Bossdorf}, \textsc{D.~Byers}, \textsc{P.~D. Kathleen},
  \textsc{M.~Dunning}, \textsc{E.~B. Holub}, \textsc{A.~Hudson}, \textsc{V.~L.
  Corre}, \textsc{O.~Loudet}, \textsc{F.~Roux}, \textsc{N.~Warthmann},
  \textsc{D.~Weigel}, \textsc{L.~Rivero}, \textsc{R.~Scholl},
  \textsc{M.~Nordborg}, \textsc{J.~Bergelson}, and \textsc{J.~O. Borevitz},
  2010, February)The scale of population structure in Arabidopsis thaliana.
\newblock PLoS Genetics~{\bf 6}\textbf{:}\ e1000843.

\bibitem[\protect\citeauthoryear{Price, Patterson, Plenge, Weinblatt, Shadick,
  and Reich}{\textsc{Price} {\em et~al.\@}}{2006}]{Price2006}
\textsc{Price, A.~L.}, \textsc{N.~J. Patterson}, \textsc{R.~M. Plenge},
  \textsc{M.~E. Weinblatt}, \textsc{N.~A. Shadick}, and \textsc{D.~Reich},
  2006\ \ Principal components analysis corrects for stratification in
  genome-wide association studies.
\newblock Nature Genetics~{\bf 38}\textbf{:}\ 904-- 909.

\bibitem[\protect\citeauthoryear{Pritchard, Stephens, and
  Donnelly}{\textsc{Pritchard} {\em et~al.\@}}{2000}]{Pritchard2000}
\textsc{Pritchard, J.~K.}, \textsc{M.~Stephens}, and \textsc{P.~Donnelly},
  2000\ \ Inference of Population Structure Using Multilocus Genotype Data.
\newblock Genetics~{\bf 155}\textbf{:}\ 945--959.

\bibitem[\protect\citeauthoryear{Redden, Divers, Vaughan, Tiwari, Beasley,
  Fern\'{a}ndez, Kimberly, Feng, Padilla, Liu, Miller, and
  Allison}{\textsc{Redden} {\em et~al.\@}}{2006}]{Redden2006}
\textsc{Redden, D.~T.}, \textsc{J.~Divers}, \textsc{L.~K. Vaughan},
  \textsc{H.~K. Tiwari}, \textsc{T.~M. Beasley}, \textsc{J.~R. Fern\'{a}ndez},
  \textsc{R.~P. Kimberly}, \textsc{R.~Feng}, \textsc{M.~A. Padilla},
  \textsc{N.~Liu}, \textsc{M.~B. Miller}, and \textsc{D.~B. Allison}, 2006,
  08)Regional Admixture Mapping and Structured Association Testing: Conceptual
  Unification and an Extensible General Linear Model.
\newblock PLoS Genet~{\em 2\/}(8)\textbf{:}\ e137.

\bibitem[\protect\citeauthoryear{Redon, Ishikawa, Fitch, Feuk, Perry, Andrews,
  Fiegler, Shapero, Carson, Chen, Cho, Dallaire, Freeman, Gonzalez, Gratacos,
  Huang, Kalaitzopoulos, Komura, MacDonald, Marshall, Mei, Montgomery,
  Nishimura, Okamura, Shen, Somerville, Tchinda, Valsesia, Woodwark, Yang,
  Zhang, Zerjal, Zhang, Armengol, Conrad, Estivill, Tyler-Smith, Carter,
  Aburatani, Lee, Jones, Scherer, and Hurles}{\textsc{Redon} {\em
  et~al.\@}}{2006}]{Redon2006}
\textsc{Redon, R.}, \textsc{S.~Ishikawa}, \textsc{K.~R. Fitch},
  \textsc{L.~Feuk}, \textsc{G.~H. Perry}, \textsc{T.~D. Andrews},
  \textsc{H.~Fiegler}, \textsc{M.~H. Shapero}, \textsc{A.~R. Carson},
  \textsc{W.~Chen}, \textsc{E.~K. Cho}, \textsc{S.~Dallaire}, \textsc{J.~L.
  Freeman}, \textsc{J.~R. Gonzalez}, \textsc{M.~Gratacos}, \textsc{J.~Huang},
  \textsc{D.~Kalaitzopoulos}, \textsc{D.~Komura}, \textsc{J.~R. MacDonald},
  \textsc{C.~R. Marshall}, \textsc{R.~Mei}, \textsc{L.~Montgomery},
  \textsc{K.~Nishimura}, \textsc{K.~Okamura}, \textsc{F.~Shen}, \textsc{M.~J.
  Somerville}, \textsc{J.~Tchinda}, \textsc{A.~Valsesia}, \textsc{C.~Woodwark},
  \textsc{F.~Yang}, \textsc{J.~Zhang}, \textsc{T.~Zerjal}, \textsc{J.~Zhang},
  \textsc{L.~Armengol}, \textsc{D.~F. Conrad}, \textsc{X.~Estivill},
  \textsc{C.~Tyler-Smith}, \textsc{N.~P. Carter}, \textsc{H.~Aburatani},
  \textsc{C.~Lee}, \textsc{K.~W. Jones}, \textsc{S.~W. Scherer}, and
  \textsc{M.~E. Hurles}, 2006\ \ Global variation in copy number in the human
  genome.
\newblock Nature~{\bf 444}\textbf{:}\ 444--454.

\bibitem[\protect\citeauthoryear{Rosenberg, Pritchard, Weber, Cann, Kidd,
  Zhivotovsky, and Feldman}{\textsc{Rosenberg} {\em
  et~al.\@}}{2002}]{Rosenberg2002}
\textsc{Rosenberg, N.}, \textsc{J.~Pritchard}, \textsc{J.~Weber},
  \textsc{H.~Cann}, \textsc{K.~Kidd}, \textsc{L.~Zhivotovsky}, and
  \textsc{M.~Feldman}, 2002\ \ The genetic structure of human populations.
\newblock Science~{\bf 298}\textbf{:}\ 2381--2385.

\bibitem[\protect\citeauthoryear{Schuster}{\textsc{Schuster}}{2008}]{Schuster2%
008}
\textsc{Schuster, S.~C.}, 2008\ \ Next-generation sequencing transforms today's
  biology.
\newblock Nature Methods~{\bf 5}\textbf{:}\ 16--18.

\bibitem[\protect\citeauthoryear{Sladek, Rocheleau, Rung, Dina, Shen, Serre,
  Boutin, Vincent, Belisle, Hadjadj, Balkau, Heude, Charpentier, Hudson,
  Montpetit, Pshezhetsky, Prentki, Posner, Balding, Meyre, Polychronakos, and
  Froguel}{\textsc{Sladek} {\em et~al.\@}}{2007}]{Sladek2007}
\textsc{Sladek, R.}, \textsc{G.~Rocheleau}, \textsc{J.~Rung}, \textsc{C.~Dina},
  \textsc{L.~Shen}, \textsc{D.~Serre}, \textsc{P.~Boutin}, \textsc{D.~Vincent},
  \textsc{A.~Belisle}, \textsc{S.~Hadjadj}, \textsc{B.~Balkau},
  \textsc{B.~Heude}, \textsc{G.~Charpentier}, \textsc{T.~Hudson},
  \textsc{A.~Montpetit}, \textsc{A.~Pshezhetsky}, \textsc{M.~Prentki},
  \textsc{B.~Posner}, \textsc{D.~Balding}, \textsc{D.~Meyre},
  \textsc{C.~Polychronakos}, and \textsc{P.~Froguel}, 2007\ \ A genome-wide
  association study identifies novel risk loci for type 2 diabetes.
\newblock Nature~{\bf 445}\textbf{:}\ 881--885.

\bibitem[\protect\citeauthoryear{Sokal, Oden, and Wilson}{\textsc{Sokal} {\em
  et~al.\@}}{1991}]{Sokal1991}
\textsc{Sokal, R.~R.}, \textsc{N.~L. Oden}, and \textsc{C.~Wilson}, 1991\ \
  Genetic evidence for the spread of agriculture in Europe by demic diffusion.
\newblock Nature~{\bf 351}\textbf{:}\ 143--145.

\bibitem[\protect\citeauthoryear{Stevens, Heckenberg, Roberson, Baugher,
  Downey, and Pevsner}{\textsc{Stevens} {\em et~al.\@}}{2011}]{Stevens2011}
\textsc{Stevens, E.~L.}, \textsc{G.~Heckenberg}, \textsc{E.~D.~O. Roberson},
  \textsc{J.~D. Baugher}, \textsc{T.~J. Downey}, and \textsc{J.~Pevsner}, 2011\
  \ Inference of Relationships in Population Data Using Identity-by-Descent and
  Identity-by-State.
\newblock PLoS Genet~{\bf 7}\textbf{:}\ e1002287.

\bibitem[\protect\citeauthoryear{Thornton and Andolfatto}{\textsc{Thornton} and
  \textsc{Andolfatto}}{2006}]{Thornton2006}
\textsc{Thornton, K.} and \textsc{P.~Andolfatto}, 2006\ \ Approximate Bayesian
  Inference Reveals Evidence for a Recent, Severe Bottleneck in a Netherlands
  Population of Drosophila melanogaster.
\newblock Genetics~{\bf 172}\textbf{:}\ 1607--1619.

\bibitem[\protect\citeauthoryear{Tiwari, Barnholtz-Sloan, Wineinger, Padilla,
  Vaughan, and Allison}{\textsc{Tiwari} {\em et~al.\@}}{2008}]{Tiwari2008}
\textsc{Tiwari, H.}, \textsc{J.~Barnholtz-Sloan}, \textsc{N.~Wineinger},
  \textsc{M.~Padilla}, \textsc{L.~Vaughan}, and \textsc{D.~Allison}, 2008\ \
  Review and Evaluation of Methods Correcting for Population Stratification
  with a Focus on Underlying Statistical Principles.
\newblock Hum Hered~{\bf 66}\textbf{:}\ 67--86.

\bibitem[\protect\citeauthoryear{Waples and Gaggiotti}{\textsc{Waples} and
  \textsc{Gaggiotti}}{2006}]{Waples2006}
\textsc{Waples, R.~S.} and \textsc{O.~Gaggiotti}, 2006\ \ What is a population?
  An empirical evaluation of some genetic methods for identifying the number of
  gene pools and their degree of connectivity.
\newblock Molecular Ecology~{\bf 15}\textbf{:}\ 1419–--1439.

\bibitem[\protect\citeauthoryear{Wasserman}{\textsc{Wasserman}}{2005}]{Wasserm%
an2005}
\textsc{Wasserman, L.}, 2005\ \ {\em All of Nonparametric Statistics}.
\newblock Springer.

\bibitem[\protect\citeauthoryear{Wright}{\textsc{Wright}}{1984}]{Wright1984}
\textsc{Wright, S.}, 1984)\ \ {\em Evolution and the genetics of populations}.
\newblock Chicago: University of Chicago Press.

\end{thebibliography}

[\textbf{terms for Glossary:}
Haploid: Organisms can have multiple copies of each chromosome that makes up their DNA. Haploid organisms, such as Bacteria and Archaea, have only one copy.

Diploid: Organisms can have multiple copies of each chromosome that makes up their DNA. Diploid organisms, such as higher animals, have two copies, one inherited from each parent.

Recombination: When an organism reproduces, it must make copies of its DNA. Sexually reproducing organisms take DNA from both parents `recombine' them in a new combination in the offspring.  Non-sexual species can also recombine their DNA but this typically does not occur at every reproduction.

Ancestral Mapping:  The identification of the genetic relationship (parents, grandparents, cousin, etc) between individuals. This is typically via a tree, although it is possible (via inbreeding) that individuals will appear multiple times in an ancestral map, e.g. if two cousins had children their grandparents would appear via both lineages.

Generative modelling: A fully specified mathematical or computer model designed to emulate a physical process, from which data can be simulated.

Statistical modelling: A statistical description of the relationship between data.  This may be associated with an underlying generative model, but does not have to be.

Coalescent, (Multilocus coalescent models): The coalescent is a tree model for the evolution of a fixed population in the absence of selective forces.  Each individual in the sample is represented as a lineage, which are each merged back in time at constant rate with each other lineage.  This process is repeated until only a single lineage remains.  In multilocus coalescent models, a number of genetic locations are each described by a coalescent.

Statistical phylogeography: A statistical approach to phylogenetic modelling (that is, representing relationships between individuals via trees) which accounts for geographical features such as sampling locations and physical barriers to population movement.

Genetic drift: The change in SNP frequency that occurs in separated populations is referred to as genetic drift. This occurs because the actual SNPs present from one generation to the next is determined stochastically.  This effect decreases with the effective population size.

Approximate Bayesian Computation: A statistical inference procedure for inferring model parameters based only on simulating from a model, without the need to compute the likelihood (i.e. the probability of the data given the parameters).

Principal Component Analysis: A linear transformation of multi-dimensional data in which the data are rotated into perpendicular axes of variation, which are sorted by the magnitude of the variation.  This can be achieved using simple linear algebra techniques and is therefore a fast method.

Demic diffusion: A genetic model in which individuals may be located in a number of geographically located populations or `demes'.  Within each deme, individuals mate at random, and migration occurs along connections between demes at a particular rate (i.e. individuals `diffuse').

Exchangeability: A statistical concept in which a number of random variables are sampled from some probability distribution, but this probability distribution does not depend on the order in which the variables are sampled. For example, if the numbers 1 through 5 are sampled without replacement, then the result is exchangeable but not independent.

Ancestral Recombination Graph: The ancestral recombination graph is a model for how individuals in a sample would relate to each other if they a) were reproducing in a closed population without selection, and b) may have more than a single parent (i.e. they are undergoing recombination). The model is statistical in that it describes only the relationship between individuals actually sampled in the population and not other individuals present in that population but who were not sampled.

Linkage disequilibrium: A genetic sample in linkage equilibrium when each SNP is independent and therefore found (on diploid organisms) on the two copies of the chromosome in the proportions expected from the population level frequency.  Linkage disequilibrium is a deviation from these proportions due to correlations on the genome caused by recombination having occurred sufficiently frequently for SNPs to be correlated.

Haplotype: A haplotype is a single copy of the set of chromosomes for an individual.  Diploid individuals consist of two haplotypes.  Specifying a haplotype rather than a genotype involves knowing which mutations appear together on the same genomic copy.

Hidden Markov Model: A Hidden Markov Model is a statistical model decomposing the probability of the data $x_i$ at locations $i$ into two parts using a set of hidden variables $y_i$.  The probability of $x_i$ depends only on $y_i$ and global parameters, i.e. it is conditionally independent of $x_j$ and $y_j$ for $j \ne i$.  The probability of $y_i$ depends on only on the previous $y_{i-1}$ but is conditionally independent of both $x_j$ (for all $j$) and $y_j$ (for $j \ne i-1$).

MCMC: Markov Chain Monte Carlo, a statistical inference technique to find the probability of the parameters given some data when this cannot be directly computed, but we can access the probability of the data given the parameters.  Parameter values are sampled, and previous `good' parameters can be used to obtain a new `good' sample.

Greedy optimisation: An optimisation method looking to find the maximum value of a function is greedy if it tries to obtain the best possible increase of the function at every iteration. It therefore cannot go down from a locally maximal value to find the globally maximal value.

HGDP: Human Genome Diversity Panel, a genetic dataset consisting of 660918 SNPs in 1043 individuals from 51 populations around the world.  See \url{http://hagsc.org/hgdp/index.html} for a description of the data and access information.

Heuristic model: A model which is not based on any theory.

Admixture: A population genetics term for the mixing of distinct populations that have been separated by genetic drift.

Parametric: A distribution that can be specified by a finite set of parameters is parametric.  For example, the mean and standard deviation are parameters of the Normal distribution.

Non-parametric: A distribution that requires a potentially infinite amount of parameters to fully specify it is non-parametric.  This is never infinite in practise -- instead, the number of parameters can grow with the number of data points observed. Because these models are very flexible, they are often claimed to make no assumptions about the data which is not quite true - the choice of how parameters describe the data can have a large effect on the inferred distribution.

Isolation-by-distance: populations undergo genetic drift separately only if there is some barrier to gene flow preventing changes in one population from appearing in the other. Isolation-by-distance models use some measure of the geographical separation between populations to describe the size of the genetic barrier, usually in terms of a migration rate that decreases with distance.
]

\end{document}